\documentclass[12pt]{article}
\usepackage[usenames]{color}
\usepackage{graphicx} 
\usepackage{amsmath}
\usepackage{amssymb}
\usepackage{amsthm}

\usepackage{hyperref}
\hypersetup{
setpagesize=false,
 bookmarksnumbered=true,%
 bookmarksopen=true,%
 colorlinks=true,%
 linkcolor=blue,
 citecolor=blue,
}
\usepackage{bm}
\usepackage{cite}
\usepackage{framed,color}
\definecolor{shadecolor}{gray}{0.95}
\usepackage{fancybox}
\usepackage{ulem}
\definecolor{shadecolor}{gray}{0.95}
\usepackage{afterpage}
\usepackage{multirow}
\numberwithin{equation}{section}

\newcommand{\vev}[1]{\left\langle #1 \right\rangle}
\newcommand{\der}{\partial}

\newcommand{\ie}{{\it i.e.}}
\newcommand{\eg}{{\it e.g.}}

\begin{document}
\begin{flushright}
\end{flushright}
\begin{center}
%
%
    {\LARGE\bf  Early Dark Energy from a Higher-dimensional Gauge Theory}
\vskip 1.4cm
{\large  
Kentaro Kojima$^{a,}$\footnote{E-mail:~kojima@artsci.kyushu-u.ac.jp}
  and 
Yuri Okubo$^{b,}$\footnote{E-mail:~okubo.yuri@phys.kyushu-u.ac.jp}%
}\\ \vskip .5cm
{\it
$^a$ Faculty of Arts and Science, Kyushu University, Fukuoka 819-0395, Japan\\%
$^b$ Graduate School of Science, Kyushu University, Fukuoka 819-0395, Japan
}\\

\vskip 1.5cm
\begin{abstract}
    The Hubble constant estimated from the CMB measurements shows
    large disagreement with the locally measured value.  This
    inconsistency is called the Hubble tension and is vastly studied
    in recent years.  Early Dark Energy (EDE) gives a few percent
    contribution to the total energy density of the universe only at
    an epoch before the recombination, and it is considered as a
    promising solution to the tension.  A simple realization of EDE is
    given by dynamics of a scalar field, called the EDE scalar, and
    models including the EDE scalar are extensively studied in the
    literature.  In this paper, we present a novel EDE scenario based
    on higher-dimensional gauge theories.  An extra component of gauge
    fields associated with a compact extra dimension behaves as the
    EDE scalar at low-energy and has a periodic potential, which has a
    similar form as potentials for pseudo Nambu-Goldstone bosons
    (PNGB).  In a five-dimensional $U(1)$ gauge theory, we show that a
    scalar field that originates from the gauge field can give EDE
    through its dynamics in a PNGB type potential with a suitable
    choice of parameters in the theory.  We focus on the scenario
    where EDE is explained by the scalar field and clarify constraints
    on the fundamental parameters of the gauge theory, such as the
    gauge coupling, the compactification scale, and the mass
    parameters for matter fields.  We also find that a sufficient
    dilution of EDE requires non-trivial relations among $U(1)$
    charges of matter fields with degenerate bulk
      masses. The required bulk matter contents and
      $U(1)$ charges can be given by hand and may be realized more
      naturally through underlying mechanisms such as flavor
      symmetries. With specific matter contents, we numerically
    solve the time evolution of the scalar field and confirm that its
    energy density behaves as an EDE.  In our scenario, the parameters
    of the gauge theory and predicted properties of EDE are related to
    each other.  Thus, the cosmological restrictions on the EDE
    properties provide insights into higher-dimensional gauge
    theories.
\end{abstract}
\end{center}
\vskip 1.0 cm
\newpage

%
\section{Introduction}
\label{sec:intro}
%
The  expansion rate of the present universe is described by the Hubble constant $H_0$, whose value  is determined through observations. 
Precise observations and studies of the cosmic microwave background~(CMB) determine the Hubble constant as $H_0=67.4\pm 0.5~ \rm{km\ s^{-1}\ Mpc^{-1}}$ by assuming the $\Lambda$CDM model of the universe~\cite{Planck:2018vyg}.
Independently  of the CMB measurements, the Hubble constant is also determined through the cosmic distance ladder. 
Recent observations of Cepheid variable stars and Type Ia supernovae as distance indicators by the SH0ES project determine the local value of the Hubble constant as $H_0=74.03\pm 1.42~ \rm{km\ s^{-1}\ Mpc^{-1}}$~\cite{Riess:2019cxk}.
There is a large discrepancy, called the Hubble tension, in the $H_0$ values inferred from CMB observations and obtained from the local measurements.  
The recent analysis shows a 5$\sigma$ difference between  them~\cite{Riess:2021jrx}. 
Since the Hubble tension arises under the assumption of the $\Lambda$CDM model, this tension may be resolved by new physics beyond the $\Lambda$CDM cosmology.

Motivated by the Hubble tension, extensions of the standard cosmology have been widely discussed in recent years~\cite{DiValentino:2021izs}.
For example, models with extra radiation have been studied~\cite{Buen-Abad:2018mas,Hryczuk:2020jhi,Vattis:2019efj,Anchordoqui:2015lqa,Pandey:2019plg,Ko:2017uyb}.
In these scenarios, additional relativistic degrees of freedom to the Standard Model (SM) of elementary particles increase the value of the Hubble constant inferred from CMB measurements.  
The Hubble tension has also been studied in models assuming extra interactions~\cite{Kumar:2016zpg,DiValentino:2017iww}, critical phenomena~\cite{Banihashemi:2018has}, and modified gravity~\cite{Khosravi:2017hfi,Odintsov:2020qzd,Yan:2019gbw,Renk:2017rzu,Dainotti:2022bzg}.
In addition, modifications of dark energy properties at late times have been studied.
For instance, the Hubble tension can be resolved by introducing phantom dark energy, whose value of the equation of state (EoS) deviates from the $\Lambda$CDM value $w=-1$ and satisfies $w<-1$~\cite{Vagnozzi:2019ezj,Alestas:2020mvb}.  
There also exist other late-time modifications of dark energy properties~\cite{DiValentino:2016hlg,DiValentino:2017zyq,Vagnozzi:2018jhn,Yang:2018qmz,DiValentino:2019jae,DiValentino:2020naf,Dainotti:2021pqg} and various theoretical approaches focusing on the discrepancy of the Hubble constant.

On the other hand, early-time modifications of the universe from the $\Lambda$CDM model have been widely discussed, motivated by the Hubble tension.  
Among them, the idea of early dark energy (EDE) is one of the promising scenarios~\cite{Karwal:2016vyq,Mortsell:2018mfj,Poulin:2018cxd,Smith:2019ihp}. 
In EDE models, the cosmic expansion is slightly modified from the $\Lambda$CDM case since an additional source of dark energy, called EDE, gives a few percent contribution to the total energy of the universe before the recombination.  
Then, the EDE density decreases faster than the matter energy density and becomes negligible, as required not to affect the cosmic  expansion after the recombination.
While the EDE scenario is restricted through the other observable quantities other than $H_0$~\cite{Hill:2020osr}, the additional energy offered by EDE increases the inferred value of $H_0$ from the CMB data and is favored for solving the Hubble tension. 
Recently, many models and analyses related to EDE have been
studied~\cite{Kamionkowski:2014zda,Poulin:2018dzj,Agrawal:2019lmo,Alexander:2019rsc,Niedermann:2019olb,Berghaus:2019cls,Sakstein:2019fmf,Ye:2020btb,Chudaykin:2020acu,Haridasu:2020pms,Vagnozzi:2021gjh,Berghaus:2022cwf}.

The EoS of EDE must depend on time.
A simple realization of the time dependence is given by the time evolution of scalar fields, as in various models of the inflation~\cite{Linde:2007fr} and quintessence~\cite{Tsujikawa:2013fta}. 
In general, kinetic and potential energies of scalar fields contribute to the energy density of the universe. Let $w_\phi$ be the value of the EoS of EDE that is supposed to be determined by a scalar field $\phi$, referred to as the EDE scalar. 
For a stationary configuration of the EDE scalar, $w_\phi\simeq -1$ is obtained.  
On the other hand, if the EDE scalar moves rapidly in its potential, then $w_\phi$ increases and deviates from $-1$.  
Thus, by using the time evolution, scalar fields can explain the time dependence of the EoS.

Several models describing the EDE scalar have been recently proposed.
The observed scalar fields, which are fundamentally described by SM of elementary particles, cannot be identified as the EDE scalar. 
Thus, if the dynamics of scalar fields give the EDE, extensions of the SM are not evaded. Therefore, the Hubble tension may also be a key to physics beyond the SM.  
A variety of models that incorporate physics beyond the SM is studied to describe the EDE scalar.  
For example, an axion-dilaton coupled system is studied~\cite{Alexander:2019rsc}, where dilaton rolls down a runaway type potential and gives energy density contributions that have EDE-like behavior.  
Also, a power-law type potential for the EDE scalar $\phi$, \ie, $V(\phi)\propto \phi^{2n}$, is valid to solve the Hubble tension as discussed in the literature~\cite{Agrawal:2019lmo,Chudaykin:2020acu}.  
If a scalar field $\phi$ oscillates in the power-law potential, the time-averaged EoS $\langle w_\phi \rangle $, related to the scalar energy density, is given by
$\langle w_\phi\rangle \simeq (n-1)/(n+1)$~\cite{Turner:1983he}.
Thus, for $n\geq 2$, the EDE density decreases faster than the matter energy density as required. 
In ref.~\cite{Chudaykin:2020acu}, by using $Planck$ 2018+SPTPol (TT \& EE)+SPTLensing+ $S_8$ prior datasets, the
Hubble constant inferred by CMB data is shown to be increased as $H_0=73.06\pm 1.26~{\rm km~s^{-1}~Mpc^{-1}}$ for $n=3$.

This paper focuses on the EDE scenario where the EDE scalar evolves in a periodic potential, which is a typical one for pseudo Nambu-Goldstone bosons (PNGBs). 
Such periodic potentials, which we call PNGB type potentials, have been investigated and considered promising to explain EDE~\cite{Poulin:2018cxd,Smith:2019ihp}.
We discuss the EDE scalar $\phi$ with the PNGB potential
$V(\phi)\propto (1-\cos[\phi/f])^n$, where $f$ is a mass parameter and $n$ is an integer. 
In models with axion-like particles, which are a familiar realization of PNGB, potentials are often considered to be generated through non-perturbative effects.
In these cases, we naturally expect that $n=1$ contributions exist in PNGB type potentials unless there is some unknown mechanism or fine-tuning. 
On the other hand, the $n=1$ contribution is implied to vanish if EDE provides a resolution to the Hubble tension~\cite{Poulin:2018dzj}. 
Thus, it is expected to reveal the origin of the EDE scalar with the $n\geq2$ PNGB type potential.
 
In this paper, we explore the validity of five-dimensional~(5D) $U(1)$ gauge theories with a compact space as a specific origin of the EDE scalar.
Higher-dimensional gauge theories are predicted in string theories and are motivated candidates for physics beyond the SM.
In 5D gauge theories, a four-dimensional (4D) scalar field appears at low energy, which originates from the extra component of the gauge field $A_5$.
Previous studies have shown that $A_5$ is a good candidate for an inflaton field~\cite{Arkani-Hamed:2003xts} or a quintessence field~\cite{Pilo:2003gu}.
As shown below, $A_5$ is also a suitable candidate for the EDE scalar. 
One advantage of this scenario is the consequence of a particular property of the potential for $A_5$.  
Due to the gauge symmetry, the potential is completely flat at the classical level and is radiatively induced.  
The potential form is highly restricted since the physical degrees of freedom are not carried by $A_5$ but by a Wilson line phase, which is a gauge-invariant and non-local quantity.
The resultant potential has a periodicity, as seen in the PNGB potential, and we obtain the EDE scalar with a PNGB potential in 5D gauge theories. 
Moreover, we find that the PNGB potential with $n\geq 2$, required to resolve the Hubble tension, is derived by using perturbative calculations in some concrete models.  
In our scenario, as a result, the fundamental parameters in 5D gauge theories, such as the gauge coupling constant and the compactification scale, control the potential for the EDE scalar and are related to EDE properties.

This paper is organized as follows.  In Sec.~\ref{sec:pngb}, we review and study the EDE scenario provided by the EDE scalar with the PNGB type potential. 
We discuss constraints on parameters of the potential to obtain adequate EDE for resolving or relaxing the Hubble tension.
In Sec.~\ref{sec:hdg}, we introduce 5D $U(1)$ gauge theories as a specific origin of the EDE scalar. Based on the one-loop effective potential for the EDE scalar, we study constraints on the mass scales appearing in 5D gauge theories.
In Sec.~\ref{sec:u1gaugeede}, we continue to discuss the 5D gauge theories assuming concrete bulk matter contents as examples. 
In this case, the EDE scalar potential depends only on a few parameters.  
We also give numerical studies of the time evolution of the EDE scalar and its energy density parameter.
Sec.~\ref{sec:conc} is devoted to discussions and conclusions.  
We derive a formula used in our discussion in the appendix.

%
\section{Constraints on the PNGB potential for the EDE scalar field}
\label{sec:pngb}
%

This section reviews the EDE resolution to the Hubble tension implemented by dynamics of scalar fields. 
We mainly discuss the case that the EDE scalar has a PNGB type potential. 
As seen below, the potential parameters are constrained to give suitable contributions to the total energy density of the universe to resolve the Hubble tension.

In the following, we assume the homogeneous, isotropic, and flat spacetime, described by the Friedmann-Lema\^itre-Robertson-Walker (FLRW) metric, as an approximation of the universe. 
The cosmic expansion is parametrized by the scale factor $a(t)$, which is the function of time $t$.
We also use the redshift $z(t)=a(t_0)/a(t)-1$, where $t_0$ is the present time. 
The Hubble parameter is given by $H(t)=\dot a(t)/a(t)$, where the dot denotes the time derivative. 
The Hubble constant $H_0$ is the present value of the Hubble parameter, \ie, $H_0=H(t_0)$.

The Hubble parameter $H(t)$ describes the expansion rate of the universe and is determined by the energy density of the universe $\rho(z)$, which we define as the function of the redshift. 
In the $\Lambda$CDM model, $\rho(z)$ consists of contributions from the non-relativistic matter $\rho_m(z)$, radiation $\rho_r(z)$, and cosmological constant $\rho_{\Lambda,0}$.  
In addition to these $\Lambda$CDM contributions, we suppose that $\rho(z)$ contains a non-negligible contribution $\rho_\phi(z)$, related to the EDE scalar. 
Thus, the total energy density is written as
\begin{align}
 \rho(z) =\rho_m(z)+\rho_r(z)+\rho_{\Lambda,0}+\rho_\phi(z).
\end{align}
We hereafter use 
$\rho_{m,0}\equiv \rho_{m}(0)$ and $\rho_{r,0}\equiv \rho_{r}(0)$.

The time evolution of the scale factor is determined by the following Friedmann equation:
\begin{align}\label{eqn:fried}
  3M_{\rm P}^2 H^2
  =\rho(z)
  =\rho_{m,0}(1+z)^3+\rho_{r,0}(1+z)^4+\rho_{\Lambda,0}+\rho_\phi(z),
\end{align}
where $M_{\rm P}=(8\pi G)^{-1/2}\simeq 2.4\times 10^{18}$~GeV is the
reduced Planck mass, and $G$ is the gravitational constant. In the $\Lambda$CDM model, we can rewrite Eq.~\eqref{eqn:fried} as
\begin{align}
  H
  =    \bar{H}_0    \sqrt{\Omega_{m,0}h^2(1+z)^3
  +    \Omega_{r,0}h^2(1+z)^4    +    \Omega_{\rm \Lambda,0}h^2    },
    \label{eqn:homega}
\end{align}
where $\Omega_{x,0}\equiv\rho_{x,0}/\rho(0)$ ($x=m,r,\Lambda$) are the
energy density parameters, and $h$ is $H_0$ in units of
$\bar{H}_0=100\ {\rm km~s^{-1}~Mpc^{-1}}$.  
By definition, the density parameters satisfy
\begin{align}\label{eqn:omegasum}
  \Omega_{m,0}+\Omega_{r,0}+\Omega_{\Lambda,0}=1, \qquad
  0\leq \Omega_{x,0} \leq 1, \quad (x=m,r,\Lambda).
\end{align}
Thus, the parameters $\Omega_{x,0}$ and $h$ determine the time evolution of $H(z)$ in the $\Lambda$CDM model.

The cosmological parameters can be determined by CMB observations.
While the curvature is also determined by the observations, we ignore it for simplicity.  
The energy density parameter of the photon $\Omega_{\gamma,0}$ can be obtained from the CMB averaged temperature $T_0$ as $\Omega_{\gamma,0}h^2=2.47\times 10^{-5}$ for $T_0=2.725~{\rm K}$\cite{Mather:1998gm}.  
We can include the neutrino contribution by using the relation $\Omega_{r,0}=(1+0.227N_{\rm eff})\Omega_{\gamma,0}$, where $N_{\rm eff}$ is the relativistic degrees of freedom of neutrinos~\cite{Lesgourgues:2006nd}, and we get $\Omega_{r,0}h^2=4.18\times 10^{-5}$ from the SM prediction $N_{\rm eff}=3.044$~\cite{Akita:2020szl}.  
The CMB power spectrum determines values of $\Omega_{m,0}$ and the baryon matter density parameter $\Omega_{b,0}$.
The ratio of even-numbered to odd-numbered peak amplitudes depends on $\Omega_{b,0}h^2$, and the ratio of the first peak to the other peaks depends on $\Omega_{m,0}h^2$.
For example, $Planck$ project gives $\Omega_{b,0}h^2=0.0224\pm 0.0001$ and $\Omega_{m,0} h^2= 0.143\pm0.001$~\cite{Planck:2018vyg}.  
From these results, the energy densities of matter and radiation coincident at $z_{\rm eq}\approx 3400$.

The angular size of the sound horizon $\theta_{*}$ at the CMB last-scattering is an observable quantity, which can be determined by the location of the first peak in the CMB power spectra. 
The parameter $\theta_{*}$ is expressed by using the comoving distance $D(z_*)$ and the sound horizon $r_{s,*}$ as
\begin{align}
    \theta_{*}
    \equiv
    \frac{r_{s,*}}{D(z_{*})},
    \label{eqn:angtheta}
\end{align}
where $z_*\approx 1090$ is the redshift at the last-scattering surface. 
The quantities $D(z_*)$ and $r_{s,*}$ depend on the Hubble parameter as
\begin{gather}
    D(z_*)
    =\int_0^{z_*}dz \frac{1}{H},
    \qquad
    r_{s,*}
    =
    \int_{z_*}^\infty dz \frac{c_s(z)}{H},
    \label{eqn:dzandhorizon}
\end{gather}
where 
\begin{align}
    c_s(z)
    =
    \left(1+\frac{3\Omega_{b,0}h^2}{4\Omega_{\gamma,0}h^2(1+z)} \right)^{-1/2},
\end{align}
is the sound speed in the baryon-photon fluid. From Eq.~\eqref{eqn:dzandhorizon}, we see that $D(z_*)$ and $r_{s,*}$ are determined by the physics after and before the recombination, respectively.  
For $z \geq z_*$, we can ignore the dark energy density parameter in Eq.~\eqref{eqn:homega}. 
Thus, using the values of $\Omega_{m,0} h^2$ and $\Omega_{r,0} h^2$ shown above, we get $r_{s,*}=0.481 \bar{H}_0^{-1}\approx 144~{\rm Mpc}$. 
On the other hand, for $z \leq z_*$, the dark energy density parameter cannot be neglected. 
In Eq.~\eqref{eqn:dzandhorizon}, we can eliminate $\Omega_{\Lambda,0}$ with the help of Eq.~\eqref{eqn:omegasum} and substitute the determined values for $\Omega_{m,0}h^2$ and $\Omega_{r,0}h^2$.  
Then, $D(z_*)$ in Eq.~\eqref{eqn:dzandhorizon} becomes a function of only $h$.  
We find an approximate relation $D(z_*)\approx 3.2/H_0$, and thus Eq.~\eqref{eqn:angtheta} implies $H_0\approx 3.2\theta_*/r_{s,*}$.  
Therefore, for given $\theta_*$ and $r_{s,*}$, we can estimate $H_0$.  
A more precise analysis has been performed using Markov Chain Monte Carlo simulations and gives $H_0=67.4\pm 0.5~ \rm{km\ s^{-1}\ Mpc^{-1}}$~\cite{Planck:2018vyg}.

The approximate relation $H_0\approx 3.2\theta_*/r_{s,*}$ implies that if $r_{s,*}$ is decreased, an inferred value of $H_0$ obtained from the CMB observations is increased.  
In EDE scenario, an additional contribution to the energy density, \ie, EDE, is supposed before the recombination. 
The EDE decreases $r_{s,*}$ and thus increases $H_0$. 
The scalar energy density $\rho_{\phi}(z)$ in Eq.~\eqref{eqn:fried} can be identified as the EDE.  
To be $D(z_*)$ unchanged, $\rho_{\phi}(z)/\rho(z) \simeq 0$ should be satisfied for $z<z_*$.

Several possibilities and models related to the origins of the EDE have recently been studied~\cite{DiValentino:2021izs}.  
In the following, we restrict our attention to the cases where the EDE appears through the dynamics of scalar fields in scalar potentials. 
We denote a homogeneous and canonical scalar field by $\phi(t)$, whose dynamics are determined by the following equation of motion:
\begin{align}\label{phi_EOM}
 \ddot \phi(t)+3H(t)\dot \phi(t)+{dV(\phi)\over d\phi}=0,
\end{align}
where $V(\phi)$ is a scalar potential. The energy density $\rho_\phi$ and the pressure $p_\phi$ of $\phi$ are given by
\begin{align}
    \rho_\phi&={1\over 2}\dot \phi^2+V(\phi), 
    \qquad 
    p_\phi={1\over 2}\dot \phi^2-V(\phi).
\end{align}

The time evolution of the energy density of the scalar field is tightly connected to the EoS, which is denoted by
$w_\phi=p_\phi/\rho_\phi$.  
If the potential contribution dominates in the energy density as $\rho_\phi\simeq V(\phi)$ and $w_\phi\simeq -1$, the energy density $\rho_\phi$ is approximately constant against expansion of the universe, like $\rho_{\Lambda,0}$. 
On the other hand, if the kinetic energy dominates the energy density as $\rho_\phi\simeq \dot \phi^2/2$ and $w_\phi\simeq 1$, $\rho_\phi$ rapidly decreases as the universe expands.

If $\rho_\phi$ gives the suitable EDE to resolve the Hubble tension, the time dependence of $w_\phi$ is restricted.  
Let us discuss the expected behavior of $w_\phi$ and the EDE scalar.  
First, $w_\phi$ varies around a specific epoch in the history of the universe. 
We denote the redshift and the time at the epoch by $z_{\rm EDE}$ and $t_{\rm EDE}$.  
The precise value of $z_{\rm EDE}$ is model dependent, and $z_*< z_{\rm EDE} \leq 10^6$ is suggested in the analysis~\cite{Poulin:2018dzj}.  
Let $\phi_{\rm min}$ be a value of the scalar $\phi$ that minimizes the potential $V(\phi)$. 
For $z\gg z_{\rm EDE}$, to obtain the time dependence of $w_\phi$, $\phi$ must deviate from $\phi_{\rm min}$ and is frozen or slowly rolling in its potential due to the Hubble friction in Eq.~\eqref{phi_EOM}. 
We denote the value of the scalar for $z\gg z_{\rm EDE}$ by
$\phi_{\rm ini}$.  
At this epoch, the energy density and the EoS satisfy $\rho_\phi\simeq V(\phi_{\rm ini})\equiv V_{\rm ini}$, and
$w_\phi\simeq -1$ holds as the EoS of dark energy. 
Since $\rho_r$ and $\rho_m$ dominate the energy density of the universe, the contribution from $\rho_\phi$ is negligible in the universe for $z\gg z_{\rm EDE}$. 
As the universe expands, $\rho_r$ and $\rho_m$ decrease, and then $\rho_\phi$ starts to give a few percent contributions to the total energy density of the universe for $z\simeq z_{\rm EDE}$. 
Simultaneously, the scalar field starts to roll down its potential quickly, which is triggered when the Hubble friction in Eq.~\eqref{phi_EOM} is sufficiently weakened. 
Thus, around $z_{\rm EDE}$, the value of the EDE scalar rapidly departs from $\phi_{\rm ini}$, and the EoS $w_\phi$ changes to a positive value. 
Finally, for $z\ll z_{\rm EDE}$, $w_\phi>0$ should be maintained to suppress $\rho_\phi$ compared to $\rho_m$.

The above discussion gives constraints on the potential for the EDE scalar $\phi$. 
As stated in Sec.~\ref{sec:intro}, we focus on the periodic potential of the following form:
\begin{align}\label{pNGBV}
  V^{(n)}(\phi)=\Lambda_{(n)}^4\left(1+\cos{\phi\over f}\right)^n, 
\end{align}
where $\Lambda_{(n)}$ and $f$ are parameters with mass dimension one, and $n\geq 1$ is an integer. 
We refer to Eq.~\eqref{pNGBV} as the PNGB potential since the above periodic potential is generally expected to appear for PNGBs, \eg, for axion-like particles.  
The PNGB potential is considered to be generated by non-perturbative effects such as instanton expansion and interpreted as the result of the spontaneous breakdown of approximate global symmetries. 
The parameter $f$ corresponds to the energy scale of the symmetry breaking.
The scale $\Lambda_{(1)}$ is the dynamical scale generated through non-perturbative effects. 
For $n\geq 2$, the potential contributions arise from higher-order instanton effects, and $\Lambda_{(n)}/\Lambda_{(1)}\ll 1$ is expected if $\Lambda_{(1)}$ is suppressed than some ultra-violet cutoff scale~\cite{Marsh:2015xka,Dimopoulos:2005ac}. 
While the potential for $\phi$ generally has the form of $V(\phi)=\sum_nV^{(n)}(\phi)$, the dominant contribution is mostly given by $n=1$ as $V(\phi)\simeq V^{(1)}(\phi)$. Nevertheless, to keep the discussion general, we consider the case with $V(\phi)\simeq V^{(n)}(\phi)$ even for $n\geq 2$ in the following.

Let us see constraints on the potential~\eqref{pNGBV} above. First, we focus on the epoch $z\gg z_{\rm EDE}$, where the EDE scalar is approximately frozen around $\phi_{\rm ini}$, which gives the initial value of the potential by $V^{(n)}(\phi_{\rm ini}) = C_0^n\Lambda_{(n)}^4$, where $C_0 \equiv1+\cos({\phi_{\rm ini}/f})$.  
Then, $C_0\sim {\cal O}(1)$ is expected without a fine-tuning of $\phi_{\rm ini}$.  
The energy density of the EDE scalar takes
$\rho_\phi\simeq V^{(n)}(\phi_{\rm ini})$ for $z \gg z_{\rm EDE}$, and $\rho_\phi$ should give a few percent contributions to the total energy density around $z_{\rm EDE}$. 
Thus, by introducing the numerical parameter $\delta={\cal O}(10^{-2})$, we obtain
\begin{align}\label{pot_scale_const1}
    \delta
    = {\rho_\phi(z_{\rm EDE})\over \rho(z_{\rm EDE})}
    \simeq{C_0^n\Lambda_{(n)}^4\over 3M_{\rm P}^2 H^2(t_{\rm EDE})},
\end{align}
which constrains the mass scale of the potential.

\begin{figure}[]
  \centering
      \includegraphics[width=9cm,clip]{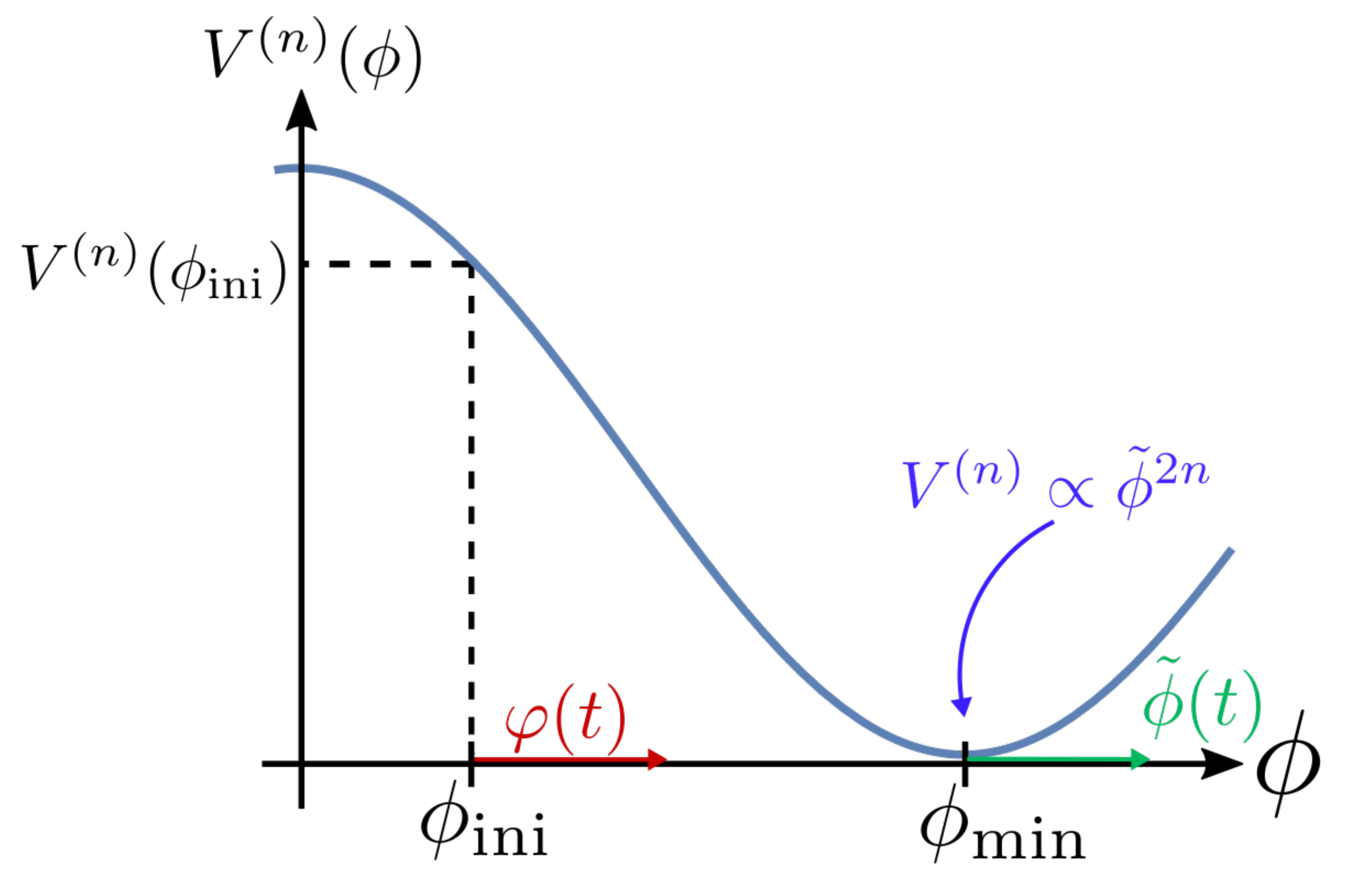}
      \caption{The periodic scalar potential $V^{(n)}(\phi)$ in Eq.~\eqref{pNGBV}. 
      The EDE scalar $\phi(t)$ is parametrized as $\phi(t)=\phi_{\rm ini} + \varphi(t)$ for $z\simeq z_{\rm EDE}$ and as $\phi(t)=\phi_{\rm min} + \tilde \phi(t)$ for $z\ll z_{\rm EDE}$. 
      Around a minimum, the potential is approximately proportional to $\tilde \phi^{2n}$.}
  \bigskip
\label{fig:Pngb}
\end{figure}

Next, we consider the dynamics of the EDE scalar around $z_{\rm EDE}$, where the scalar starts to change its value from $\phi_{\rm ini}$. 
Let us parametrize $\phi=\phi_{\rm ini}+\varphi$ as shown in Fig.~\ref{fig:Pngb} and assume $|\varphi/f|\ll 1$ for $z\simeq z_{\rm EDE}$. 
Then, Eq.~\eqref{phi_EOM} determines the evolution of $\varphi$ as
\begin{align}\label{eomvarphi1}
    3H {\dot \varphi\over f}
  \simeq -{1\over f}{d\over d\phi}V^{(n)}(\phi_{\rm ini}+\varphi)
  \simeq
    nC_0^{n-1}\frac{\Lambda_{(n)}^4}{f^{2}}\left[
    -\left\{(2-C_0)n-1\right\}
    {\varphi\over f}
    +  \sin{\phi_{\rm ini}\over f} \right],
\end{align}
where we have approximately neglected the first term in
Eq.~\eqref{phi_EOM}.  
In addition, we only take the leading order of $\varphi/f$ in the above last equation. 
If we parametrize the Hubble parameter as $H(t)=b/t$, where $b=2/3$ ($1/2$) in the matter-dominated (radiation-dominated) era, the solution of Eq.~\eqref{eomvarphi1} is written as follows:
\begin{align}\label{solphi}
  {|\varphi|\over f}\simeq &
  {\sin({\phi_{\rm ini}/f}) \over
  |(2-C_0)n-1|
  }
  \left[
  \exp\left(
  {nC_0^{n-1}|(2-C_0)n-1|\Lambda_{(n)}^4\over 6bf^2}t^2  \right)
  -1
  \right].
\end{align}
This solution implies the following relation: 
\begin{align}\label{htede1}
  H^2(t_{\rm EDE})\simeq  {bnC_0^{n-1}|(2-C_0)n-1|
  \over 6}\Lambda_{(n)}^4f^{-2}, 
\end{align}
which also constraints the parameters of the potential.

From Eqs.~\eqref{pot_scale_const1} and~\eqref{htede1}, we obtain
\begin{align}\label{potmasscon1}
  \Lambda_{(n)}\simeq
  \left({3\delta\over C_0^n}\right)^{1/4} \sqrt{H(t_{\rm EDE})M_{\rm P}},
  \qquad
  f\simeq \left({3\delta\over C_0^n\tilde C}\right)^{1/2} M_{\rm P},
\end{align}
where $ \tilde C\equiv 6 (bNC_0^{n-1}|(2-C_0)n-1|)^{-1}$ is a dimensionless parameter. 
Without fine-tuning of the parameters, $C_0$ and $\tilde C$ in Eq.~\eqref{potmasscon1} are expected to be ${\cal O}(1)$. In addition, $H(t_{\rm EDE})\sim 10^{-28}~{\rm eV}$ is satisfied for $z_{\rm EDE}\sim 4000$, which is an example and a typical value. 
Thus, the parameters in the PNGB potential in Eq.~\eqref{pNGBV} should satisfy $\Lambda_{(n)}\sim 1$~eV and $f\sim M_{\rm P}$ in the EDE scenario. 
This result is not sensitive to a precise value of $z_{\rm EDE}$.

Finally, we consider the dynamics of the EDE scalar for $z < z_{\rm EDE}$. 
The scalar begins to evolve to a minimum of the potential in Eq.~\eqref{pNGBV} around $z_{\rm EDE}$ and then rapidly oscillates around the minimum. 
Let $\tilde \phi$ be a fluctuation of the scalar from the minimum as shown in Fig.~\ref{fig:Pngb}. 
For $z\ll z_{\rm EDE}$, Eq.~\eqref{pNGBV} implies that $\tilde \phi$ oscillates in the following polynomial potential:
\begin{align}\label{pol_pot_1}
  V^{(n)}(\tilde \phi)\simeq {\Lambda_{(n)}^4\over 2^n}\left({\tilde \phi\over f}\right)^{2n}.
\end{align}
The period of the oscillation is characterized by the time scale of ${\cal O}(1/H(z_{\rm EDE}))$, and the EoS of the scalar also oscillates for $z\ll z_{\rm EDE}$.  
Due to the Hubble friction term in Eq.~\eqref{phi_EOM}, the oscillation amplitude decreases with a time scale, which is much longer than the oscillation period.  
In this case, the time-averaged value of the EoS $\vev{w_{\phi}(z)}$ is approximately given by~\cite{Turner:1983he}
\begin{align}\label{avephieos}
  \vev{ w_{\phi}(z)}\simeq {n-1\over n+1}, \qquad {\rm for}\qquad z\ll z_{\rm EDE}.
\end{align}
Therefore, to decrease the energy density $\rho_\phi$ faster than $\rho_m$, $n\geq 2$ is required~\cite{Poulin:2018dzj}. In the $n=2$ case, $\vev{w_\phi(z)}\simeq 1/3$ holds during the oscillation, and $\rho_\phi$ behaves like $\rho_r$. 
Thus, the EDE scenario with the $n=2$ potential and models with extra radiation components, which modify $N_{\rm eff}$~\cite{Planck:2018vyg,Riess:2016jrr,Kreisch:2019yzn,DEramo:2018vss}, are partly degenerate in their predictions.  
A novel and the minimal setting is the $n=3$ case, which is suggested to be the most effective to the Hubble tension in the analysis in~\cite{Poulin:2018cxd}.

We summarize the discussion in this section.  We focus on the EDE scenario in which the scalar dynamics in the PNGB
potential~\eqref{pNGBV} explain the time evolution of the EDE. 
The mass parameters in Eq.~\eqref{pNGBV} are constrained as
$\Lambda_{(n)}\sim 1$~eV and $f\sim M_{\rm P}$.  
In addition, the leading potential must be $n\geq 2$ contributions to obtain a sufficient suppression of the EDE for $z\ll z_{\rm EDE}$.  
The vast hierarchy in the potential parameters implies an underlying mechanism that controls the potential.  
Ultralight axions can have potentials with hierarchical mass scales~\cite{Marsh:2015xka}, and they are considered to be a candidate for the EDE scalar~\cite{Smith:2019ihp,Poulin:2018cxd,Poulin:2018dzj}.  
However, as noted above, the PNGB potential is expected to be dominated by the $n=1$ contribution, which is inadequate for the EDE scenario. 
Thus, if an axion-like particle is identified as the EDE scalar, an unknown mechanism, which ensures that $n\geq 2$ contributions dominate the potential, is required.  
In this paper, as an alternative to the axion case, we point out that the scalar potential suitable to the EDE scenario appears in models incorporating a compact extra dimension. 
As seen below, an extra-dimensional component of gauge fields naturally has a potential adequate to the EDE scalar through quantum corrections.

%
\section{Higher-dimensional gauge field as the EDE scalar}
\label{sec:hdg}
%
As discussed in the previous section, the potential for the EDE scalar should satisfy severe constraints to solve the Hubble tension. 
Since there is no candidate for the EDE scalar among the known particles in nature, physics beyond the SM is implied to explain the EDE scalar.
Although we can introduce a scalar field and fix the scalar potential by hand, it needs extreme fine-tuning; this is because scalar potentials generally receive quantum corrections, and the potential energy tends to be comparable to an ultraviolet cutoff scale. 
Thus, we should address the problem of revealing the underlying physics of the EDE scalar from the particle physics viewpoint.

Here, we discuss the possibility that the EDE scalar naturally emerges from higher-dimensional gauge theories, which are well-motivated candidates for physics beyond the SM.  
In these theories, there are higher-dimensional components of gauge fields, which are regarded as 4D scalar fields at low energy, and may be identified as the EDE scalar. 
The gauge invariance strictly restricts their potentials, and as examined below, we can obtain the effective potential, which is adequate to EDE resolution to the Hubble tension, without fine-tuning in higher-dimensional gauge theories.

As concrete examples, let us discuss the extra-dimensional models based on a five-dimensional (5D) $U(1)$ gauge theory, where 5D spacetime consists of the Minkowski spacetime and a compact space. 
In the following, we assume for simplicity that the gauge group, which we denote by $U(1)_D$, is the gauge symmetry for a dark sector. 
Namely, the SM fields, such as baryons and leptons, are neutral under the $U(1)_D$ symmetry. 
We denote the spacetime coordinates by $x^M$ $(M=0,1,2,3,5)$; we also use $x^\mu$ ($\mu=0,1,2,3$) and $y=x^5$. 
The geometry of the compact extra space is assumed to be a circle $S^1$ or an orbifold $S^1/{\mathbb Z}_2$. In both cases, we denote the radius of $S^1$ by $R$. 
Thus, for the $S^1$ compactification, we demand the identification $y\sim y+2\pi R$.  For the $S^1/{\mathbb Z}_2$ compactification, the extra identification $y\sim -y$ is also imposed.

In the 5D gauge theory, the gauge field is denoted by
$A_M=(A_\mu,A_5)$. As matter fields for the dark sector, we introduce $N_f$ and $N_b$ flavors of 5D fermion fields $\psi_m$ $(m=1,\dots,N_f)$ and complex scalar fields $\phi_m$ $(m=1,\dots,N_b)$, respectively.  
The matter fields are assumed to be charged under the $U(1)_D$ and neutral under the SM gauge symmetry. 
The action and the Lagrangian density of the dark sector is written as follows:
\begin{gather}
  S=\int d^4 x \int_0^{2\pi R} dy ({\cal L}_g+{\cal L}_\varphi), \\
  {\cal L}_g=-{1\over 4}F_{MN}F^{MN},\quad F_{MN}=\der_MA_N-\der_NA_M, \\
  {\cal L}_\varphi
  =\sum_{m=1}^{N_f}\bar \psi_m(i\Gamma^MD_M-M_{\psi_m})\psi_m+\sum_{m=1}^{N_b}\left[|D_M\phi_m|^2-M^2_{\phi_m}|\phi_m|^2\right],
\label{lagmat1}
\end{gather}
where $M_{\varphi}$ $(\varphi=\psi_m,\phi_m)$ are bulk mass parameters and $\Gamma^M=(\gamma^\mu,i\gamma^5)$ are the 5D gamma matrices.  
The covariant derivative $D_M$ is given by $D_M=\der_M-ig_Dq_\varphi A_M$, where $g_D$ is the gauge coupling and $q_\varphi$ $(\varphi=\psi_m,\phi_m)$ is a $U(1)_D$ charge of $\varphi$. 
The Lagrangian density is invariant under the following
gauge transformation:
\begin{align}\label{gtdef1}
  A_M\to A_M+{1\over g_D}\der_M\Lambda, \quad
  \varphi\to e^{iq_\varphi \Lambda}\varphi, \quad (\varphi=\psi_m,\phi_m).
\end{align}

At an energy scale below the compactification scale $1/R$, the effective 4D description is valid, where a five-dimensional field is expanded as infinite towers of Kalzua-Klein (KK) modes that only depend on the 4D coordinates.
Except for zero-modes, whose wave functions are completely flat along the extra dimension, the KK modes have large masses of ${\cal O}(1/R)$ at the classical level.
We suppose that $1/R$ is much larger than the typical energy scale of the EDE scalar potential. 
Thus, we focus on the case where $A_5$ has a zero-mode.

The KK expansion is determined by boundary conditions, which relate the field values at $(x^\mu,y)$ and $(x^\mu,y')$ under the identification $y\sim y'$. 
The boundary conditions can be chosen such that the Lagrangian densities at $(x^\mu,y)$ and $(x^\mu,y')$ have the same value~\cite{Hebecker:2001jb}. 
To obtain an $A_5$ zero-mode, one can introduce the boundary condition $A_M(x^\mu,y+2\pi R)=A_M(x^\mu,y)$ for the $S^1$ compactification.
In this case, both $A_\mu$ and $A_5$ have zero-modes. 
On the other hand, for the $S^1/{\mathbb Z}_2$ compactification, there is an additional boundary condition associated with the identification $y\sim-y$. 
Under the parity transformation of the fifth dimension, $y\to -y$, $A_5$ changes its sign. 
Thus, in the $S^1/{\mathbb Z}_2$ case, we cannot take common boundary conditions for $A_\mu$ and $A_5$.
To obtain an $A_5$ zero-mode, we can consistently introduce the additional boundary conditions as $A_\mu(x,-y)=-A_\mu(x,y)$ and $A_5(x,-y)=A_5(x,y)$. 
These boundary conditions are related to the outer automorphism, which is the complex conjugation of the
representation of $U(1)_D$, and are called the conjugate boundary condition~\cite{Haba:2008ar,Kawamura:2011re,Abe:2016tfq}.  
In this case, while $A_5$ has its zero-mode, $A_\mu$ does not have, and the $U(1)_D$ gauge symmetry is broken.

In these cases, the $A_5$ zero-mode can be identified as the EDE scalar. 
Hereafter, we denote the $A_5$ zero-mode by $A_5^{(0)}$. 
The canonical normalization of $A_5^{(0)}$ implies $A_5\ni A_5^{(0)}/\sqrt{2\pi R}$, and the effective 4D gauge coupling $\tilde g_D$ is given by $\tilde g_D=g_D/\sqrt{2\pi R}$.

For the matter fields, boundary conditions should also be considered.
For the $S^1$ case, we impose $\varphi(x^\mu,y+2\pi R)= e^{i\eta_\varphi}\varphi(x^\mu,y)$ $(\varphi=\phi_m,\psi_m)$, where $\eta_\varphi\in {\mathbb R}$ is a parameter.
For the $S^1/{\mathbb Z}_2$ case, we introduce the additional conditions as $\phi_m(x^\mu,-y)= \phi_m^*(x^\mu,y)$ and $\psi_m(x^\mu,-y)= i\psi_m^c(x^\mu,y)$, where $c$ denotes the charge conjugation of the 4D Dirac fermion.
In both cases, the bulk mass terms for the matter fields in Eq.~\eqref{lagmat1} are invariant under the boundary conditions.\footnote{In $S^1/{\mathbb Z}_2$ case, depending on boundary conditions, we should incorporate a pair of fermions with opposite ${\mathbb Z}_2$ parities to construct invariant mass terms under the boundary conditions~\cite{Maru:2006ej}.}

Let us discuss the potential for $A_5^{(0)}$. The gauge field has a flat potential at the classical level, and an effective potential is radiatively generated.
We denote a homogeneous background value by $\langle A_5^{(0)}\rangle $. 
Note that $\langle A_5^{(0)}\rangle $ is gauge dependent since gauge transformations in Eq.~\eqref{gtdef1} change the background value.  
The Wilson line defined with a non-contractible loop along the compact extra-dimension is a gauge-invariant quantity and has physical consequences in the present setup. 
The background value of the Wilson line $W$ is written as \begin{align}
  \langle W\rangle=e^{ig_D\int_0^{2\pi R}dy \left\langle A_5\right\rangle}  
  =e^{{2\pi i R}\tilde g_D \left\langle A_5^{(0)}\right\rangle}=e^{i \theta_5}, 
\end{align}
where we have defined the phase parameter $\theta_5=2\pi R\tilde g_D \langle A_5^{(0)}\rangle $ and taken the minimum unit of $U(1)_D$ charge to be 1.  
Due to the phase property of the Wilson line, the effective potential for $\langle A_5^{(0)}\rangle$ is given by a periodic function of $\theta_5$, \ie, $V(\theta_5)=V(\theta_5+2\pi n)$ $(n\in {\mathbb Z})$.
If a matter field $\varphi$ having a bulk mass $M_\varphi$ exists, the following contribution to the effective potential $V(\theta_5)$ is generated at the one-loop level~\cite{Delgado:1998qr,Maru:2006ej}:
\begin{align}
  \Delta V_\varphi(\theta_5)
  =
  \frac{3n_\varphi M_c^4}{4\pi^2}
  \sum_{\ell=1}^\infty
  e^{-\ell M_\varphi/M_c}
  \left[
  \frac{1}{\ell^5}
  +
  \frac{1}{\ell^4}
  \frac{M_\varphi}{M_c}
  +
  \frac{1}{3\ell^3}
  \frac{M_\varphi^2}{M_c^2}
  \right]
  \cos [\ell(q_\varphi \theta_5-\eta_\varphi)],
  \label{vtheta5w}
\end{align}
where we have defined $M_c=(2\pi R)^{-1}$. 
The factor $n_\varphi$ takes $4$ $(-2)$ if $\varphi$ is a 5D fermion (complex scalar) for the $S^1$ case, whereas $n_\varphi$ takes $2$ $(-1)$ for the $S^1/{\mathbb Z}_2$ case.  
In general, there is a constant contribution to the potential, which is ignored in Eq.~\eqref{vtheta5w} and can be canceled by a constant dark energy.
Also, additional contributions to $V(\theta_5)$ can arise through 5D gravitational interactions, which may be large and change the potential structure in Eq.~\eqref{vtheta5w}. 
However, if $M_c/M_5<1$ is satisfied, where $M_5$ is the 5D Planck mass parameter, such quantum gravity corrections are safely neglected~\cite{Arkani-Hamed:2003xts}. 
Thus, we consider the case with $M_c/M_5<1$.  
The one-loop effective potential for $\theta_5$ is given by the sum of the contributions from the matter fields in the model, \ie, $V(\theta_5)=\sum_\varphi \Delta V_\varphi(\theta_5)$.

From Eq.~\eqref{vtheta5w}, if $M_\varphi/M_c > 1$ is satisfied, we find that the potential contribution is approximately written by
\begin{align}
\Delta  V_\varphi(\theta_5)\simeq 
  \frac{n_\varphi M_\varphi^2M_c^2}{4\pi^2}
  e^{-M_\varphi/M_c}
  \cos(q_\varphi \theta_5-\eta_\varphi).
  \label{effpot}
\end{align}
In this case, the potential is given by the simple cosine form as the PNGB potential in Eq.~\eqref{pNGBV}, and the exponential factor in Eq.~\eqref{effpot} strongly suppresses the energy scale of the potential compared to $\sqrt{M_\varphi M_c}$.
This suppression reflects the non-locality of the Wilson line phase degrees of freedom. 
Since the potential contributions can be seen as the result of virtual propagations of the massive matter fields along non-contractible cycles, the contribution scales like $e^{-2\pi R \ell M_\varphi}$.  
Thus, if $M_\varphi/M_c > 1$ is satisfied, we can naturally obtain a suitable EDE scale through the potential in Eq.~\eqref{effpot} with a wide parameter range of the compactification scale $M_c$.

\begin{figure}[]
  \centering
      \includegraphics[width=10cm,clip]{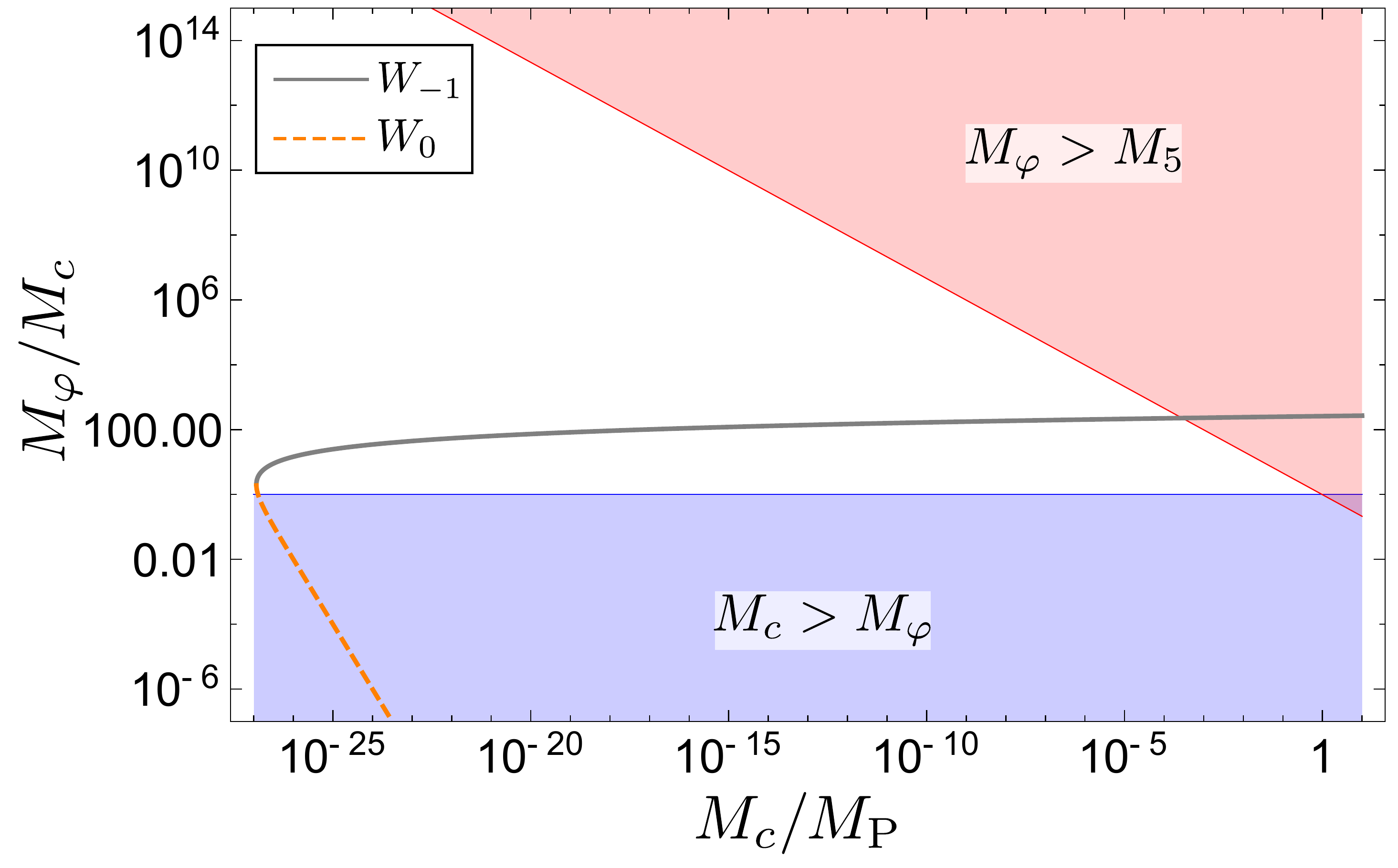}
      \caption{ The allowed range of the mass parameters in the
        scenario where the extra-dimensional gauge field is identified as the EDE scalar. 
        The blue shaded region is inconsistent with the first inequality in Eq.~\eqref{massrel}, where a PNGB-like potential~\eqref{effpot} with a suppressed energy scale suitable for the EDE scalar potential cannot be obtained. 
        The red shaded region is inconsistent with the second inequality in Eq.~\eqref{massrel}, where the gravitational corrections to the potential are not safely neglected.  
        The solid (black) and dashed (red) lines correspond to the solution of Eq.~\eqref{bulkcomp} for a given $M_c/M_{\rm P}$.}
  \bigskip
\label{fig01}
\end{figure}

Based on the above discussions, we study the constraints for the mass parameters in the scenario where the EDE scalar is the extra-dimensional gauge field. 
To obtain the PNGB-like potential~\eqref{effpot} for the EDE scalar, the condition,
\begin{align}
  M_c  <   M_\varphi   <  M_5,
	\label{masscond}
\end{align}
should be satisfied.  Since the 4D Planck mass $M_{\rm P}$ is given by the 5D Planck mass $M_5$ and the compactification scale $M_c$ as $ M_5^3=M_cM_{\rm P}^2$, we can rewrite the condition~(\ref{masscond}) as
\begin{align}
	1<\frac{M_{\varphi}}{M_c}<\left(\frac{M_{\rm P}}{M_c} \right)^{2/3}.
  \label{massrel}
\end{align}
In Fig.~\ref{fig01}, we show the allowed range of the parameters in this scenario, where the blue (red) shaded region is not consistent with the first (second) inequality in Eq.~\eqref{massrel}.

Let us examine the typical parameters that give the required energy scale for EDE. 
As discussed in Sec.~2, if the EDE scalar potential is given by Eq.~\eqref{pNGBV}, $\Lambda_{(n)}\sim 1$~eV and $f\sim M_{\rm P}$ should be satisfied. 
In the present case, a similar discussion gives the following constraints:
\begin{align}
  \frac{
  M_\varphi^2 M_c^2}{4\pi^2}e^{-M_\varphi/M_c}\sim (1~{\rm eV})^4,
  \qquad \frac{M_c}{
  \tilde{g}_D}\sim M_{\rm P},
  \label{5Dmasscon1}
\end{align}
where we have ignored $\mathcal{O}(1)$ factors such as $n_\varphi $ and $q_\varphi$.
Typical values of ${M_\varphi/M_c}$ that satisfy the constraints in Eq.~\eqref{5Dmasscon1} are given by the solutions of the following equation:
\begin{align}
	\frac{M_\varphi}{M_c}
	=
	-2W_k
	\left(
		-\frac{10^{-54}}{2}
		\left(
			\frac{M_c}{M_{\rm P}}
		\right)^{-2}	
	\right)\ ,
	\label{bulkcomp}
\end{align}
where $W_k$ $(k=-1,0)$ is the Lambert $W$ function. 
For a given $M_c/M_{\rm P}\geq 10^{-27}e/2$, where $e= 2.71828\dots$, there are two solutions of Eq.~\eqref{bulkcomp}, which we denote in Fig.~\ref{fig01} as the solid and the dashed lines. 
As seen from the figure, only the solution of the $k=-1$ case of Eq.~\eqref{bulkcomp} is consistent with the constraint in Eq.~\eqref{masscond}.

From the above analysis, we find that a wide range of the
compactification scale is allowed for the present scenario with $M_c/M_{\rm P} \lesssim 10^{-3}$. 
The ratio between the bulk mass scale for matter fields and $M_c$ should satisfy $M_\varphi/M_c\sim \mathcal{O}(100)$. 
The $\ell\geq 2$ terms in the effective potential~\eqref{effpot} are consistently neglected with the typical values of $M_\varphi/M_c$.  
Note that the second relation in Eq.~\eqref{5Dmasscon1} implies that the required value of the 4D gauge coupling constant $\tilde{g}_D$ is determined by $M_c/M_{\rm P}$, which is $\tilde{g}_{D}\lesssim 10^{-3}$ in the allowed region.
Such a small value of $\tilde{g}_D$ is consistent with the perturbative evaluation and the one-loop approximation of the effective potential in Eq.~\eqref{vtheta5w}.

In this section, we have studied the scenario where the 5D component of the gauge field $A_5$ in a 5D dark sector is identified as the EDE scalar. 
The compactification scale of the extra dimension $M_c$ is smaller than ${\cal O}(10^{-3}M_{\rm P})$.  
The effective 4D gauge coupling in the dark sector $\tilde g_D$ is predicted to be comparable to $M_c/M_{\rm P}$.
If a bulk mass for a matter field satisfies $M_\varphi/M_c\sim {\cal O}(100)$, the effective potential for $A_5$ is sufficiently suppressed to give the tiny energy scale of the EDE scalar potential without extreme fine-tuning of the parameters.
In addition to these conditions, to obtain realistic EDE potentials, EDE should decrease faster than the matter energy density in the late universe. 
As discussed in Sec.~\ref{sec:pngb}, such late-time behavior of EDE is described by $n\geq 2$ potential of Eq.~\eqref{pol_pot_1} around its minimum. 
As we will see, $A_5$ potentials also explain the late-time behavior of EDE if there are both fermions and scalars as the bulk matter fields in the dark sector.

%
\section{Explicit examples of 5D $U(1)_D$ EDE models}
\label{sec:u1gaugeede}
%
In this section, we show examples of 5D $U(1)_D$ models that give EDE. 
For simplicity of the discussion, we assume that the bulk matter fields satisfy the periodic boundary conditions, \ie, $\eta_\varphi=0$, and have a degenerate bulk mass as $M_\varphi=M$.
In this case, the effective potential for $\theta_5$ is written by 
\begin{align}\label{th5pot1}
\begin{aligned}
  V_{\rm eff}(\theta_5)=&-V_0\sum_{\varphi} n_\varphi \left[1-\cos(q_\varphi \theta_5)\right],
  \\
  V_0   &=   \frac{M^2M_c^2}{4\pi^2}e^{-M/M_c},
\end{aligned}
\end{align}
where we have added a constant contribution to the potential in order to simplify the following discussion. We hereafter refer to $\theta_5$ as the EDE phase.

If we specify the matter contents of the model, the effective potential in Eq.~\eqref{th5pot1} is determined. 
As seen in Sec.~\ref{sec:pngb}, if an EDE scalar oscillates around a minimum of $n\geq 2$ potential in Eq.~\eqref{pol_pot_1}, EDE dilutes sufficiently faster than the matter energy density in the late universe. 
Such dilution can be ensured in our setup by specific bulk matter contents.
To see this, we focus on the late-time behavior of the EDE phase, which is assumed to oscillate around $\theta_5= 0$ in the potential.
For $|\theta_5|\ll 1$, the potential is approximately written as
\begin{align}
  V_{\rm eff}(\theta_5)\simeq V_0\sum_{k=1}^\infty{(-1)^{k}\over 2k!}\left(\sum_{\varphi}n_\varphi q_\varphi^{2k}\right)
  \theta_5^{2k}.
\end{align}
Let us define
\begin{align}
  V_{\rm eff}^{(n)}(\theta_5)\equiv
  V_0\left(\sum_{\varphi}n_\varphi q_\varphi^{2n}\right)
  {(-1)^{n}\over 2n!}\theta_5^{2n}+{\cal O}(\theta_5^{2n+2}).
\end{align}
Thus, if $V_{\rm eff}(\theta_5)\simeq V_{\rm eff}^{(n)}(\theta_5)$ is
satisfied for $|\theta_5|\ll 1$, the EoS of the EDE phase is given by the same form as Eq.~\eqref{avephieos}. 
For $n\geq 2$, EDE dilutes faster than the matter energy density.

To obtain $V_{\rm eff}(\theta_5)\simeq V_{\rm eff}^{(n)}(\theta_5)$, bulk matter contents are restricted, and the parameters related to the matter contents should satisfy
\begin{align}\label{nqconst1}
  \sum_{\varphi}n_\varphi q_\varphi^{2k}=0, \qquad {\rm for} \qquad k=1,\dots,n-1.
\end{align}
For a given set of fields $\varphi$ with charge $q_\varphi$, we can regard Eq.~\eqref{nqconst1} as constraints for $n_\varphi$.
If there are equal to or more than $n$ kinds of fields with different $U(1)_D$ charges, there exist solutions of $n_\varphi$ in Eq.~\eqref{nqconst1} as shown in Appendix~\ref{App:geneff}.

We show an explicit example of models that give $V_{\rm eff}(\theta_5)\simeq V_{\rm eff}^{(n)}(\theta_5)$.
Let us express bulk scalars or fermions whose $U(1)_D$ charges are $q_i$ by $\varphi_i$ $(i=1,\dots,n)$, where $|q_i|\neq |q_j|$ for $i\neq j$. 
In addition, let $|d_i|$ be the sum of the degrees of freedom of $\varphi_i$, where $d_i<0$ ($d_i>0$) for scalars (fermions). 
In this case, the solution of the constraint~\eqref{nqconst1} is given by
\begin{align}\label{sol_nqconst1}
  {d_i\over d_j}=-{q_j^2\over q_i^2}\prod_{\substack{k=1\\k\neq i,j}}^n{q_j^2-q_k^2\over q_i^2-q_k^2},
\end{align}
for $i\neq j$. 
The effective potential is approximately written as follows:
\begin{align}\label{approx_leadingpot1}
  V_{\rm eff}(\theta_5)
  &\simeq V_{\rm eff}^{(n)}(\theta_5)
  =
  -{V_0\over (2n)!}d_{j}q_{j}^{2}\left(\prod_{\substack{k=1\\k\neq j}}^n(q_k^2-q_j^2)
  \right)\theta_5^{2n}+{\cal O}(\theta_5^{2n+2}),
\end{align}
which holds for any $j$.

For example, $V_{\rm eff}(\theta_5)\simeq V_{\rm eff}^{(2)}(\theta_5)$ is given by the bulk matter content consisting of eight complex scalars with charge $q_1$ and a 5D fermion with $q_2= 2q_1$ or $-2q_1$. 
We refer to the above case as the $n=2$ model hereafter.
Also, the $n=3$ model is defined by, \eg, the bulk matter content consisting of 15 complex scalars with charge $q_1$, a complex scalar with charge $q_2=3q_1$ or $-3q_1$, and three 5D fermions with charge $q_3=2q_1$ or $-2q_1$.\footnote{ If we assume the anti-periodic boundary condition for fermions and scalars, the right-hand side of Eq.~\eqref{th5pot1} is multiplied by $-1$ with a suitable constant shift of the potential. 
In that case, \eg, $V_{\rm eff}(\theta_5)\simeq V_{\rm eff}^{(2)}(\theta_5)$ is given by the bulk matter content consisting of two 5D fermions with charge $q_1$ and a complex scalar with charge $q_2=2q_1$ or $-2q_1$.}
The required bulk matter contents and $U(1)$ charges for $n\geq 2$ models can be given by hand and regarded as a setup. Also, a suitable matter sector may be realized more naturally through underlying mechanisms such as flavor symmetries.

With the above specific matter contents, let us discuss the time evolution of the EDE phase and its energy density.
In a 4D low-energy effective theory, $\langle A_5^{(0)} \rangle$ is a canonically normalized scalar field, which obeys the following equation of motion:
\begin{align}\label{eomth51}
  \vev{\ddot A_5^{(0)}}+3H(t) \vev{\dot A_5^{(0)}}+2\pi R\tilde g_D{dV_{\rm eff}(\theta_5)\over d\theta_5}=0.
\end{align}
Let $w_{\theta_5}$ and $\rho_{\theta_5}$ be the EoS and the energy density of the EDE phase, respectively.  
In addition, we denote the energy density parameter of $\theta_5$ by $\Omega_{\theta_5}$, which is the ratio between $\rho_{\theta_5}$ and the total energy density of the universe.
If the potential behaves as $V_{\rm eff}(\theta_5)\simeq V_{\rm eff}^{(n)}(\theta_5)$, the time average of $w_{\theta_5}$ after a fast-roll period is well approximated by the right-hand side in Eq.~(\ref{avephieos}).

For fixed matter contents and a compactification scale $M_c$, the time evolution of the EDE phase depends on $\tilde g_D$ and $M$.
The latter controls the energy scale of $V_{\rm eff}(\theta_5)$ as understood from Eq.~\eqref{th5pot1}. 
On the other hand, $\tilde g_D$ is irrelevant to the potential and affects the equation of motion~\eqref{eomth51}. 
As discussed in Sec.~\ref{sec:pngb}, $\theta_5$ is initially frozen in its potential. 
Then, the Hubble friction in Eq.~\eqref{eomth51} becomes sufficiently small, and $\theta_5$ starts to roll down the potential. 
If $\tilde g_D$ is increased (decreased) for a given potential, the effect of the Hubble friction term is diminished (enhanced).  
Thus, $\tilde g_D$ controls the time when the fast roll starts.

To examine typical dynamics of the EDE phase, we numerically solve the equation of motion of the EDE phase in the $n=2$ and $3$ models. 
We set $q_1 = 1$ and initialize the system at $t=t_0$, where
$z(t_0)=10^{42}$.
Since EDE accounts for less than several percent of the total energy of the universe, the expansion rate of the universe is almost determined by the evolution of the other ingredients.
In the following analysis, we consider a universe in the matter-dominant stage, \ie, the Hubble parameter scales as $H(t)=2/(3t)$ to simplify calculations.
For $z\lesssim z_{\rm eq}$, our universe is well approximated by this treatment.  
In addition, from the discussion given in Sec.~\ref{sec:pngb}, we see that the dynamics of the EDE phase is not sensitive to the value of $b$ under the parametrization $H(t)=b/t$.
While EDE may generally become non-negligible during the radiation-dominated period, where $H(t)=1/(2t)$ holds, or at the matter-radiation equality time, our treatment is sufficient to understand the qualitative behavior of the EDE phase.

To compare the $n=2$ and $3$ models, we choose the parameters such that the fast-roll of the EDE phase begins at almost the same time in each model. 
In addition, we set similar maximum values of the EDE density parameters for both cases. These conditions fix the values of $M/M_{\rm P}$ and $\tilde g_D$.
The parameters and initial conditions for the EDE phase used in the following analysis are shown in Table~\ref{table01}.

\begin{table}[]
\centering
\caption{The parameters and initial conditions for the EDE phase used in the numerical analysis. The predicted value of the 4D gauge coupling constant is also shown.  }
\bigskip
\begin{tabular}{c|ccccc } \hline\hline
   model & $M_c/M_{\rm P}$ & $M/M_c$ & $\tilde g_D$ &$\theta_5(t_0)$& $\dot \theta_5(t_0)$
  \\ \hline
$n=2$ & $1.0\times 10^{-6}$&$212.5$&$6.95\times 10^{-6}$&$\pi/2$&$0$\\
  $n=3$ & $1.0\times 10^{-6}$&$213.1$& $5.00\times 10^{-6}$&$\pi/2$&$0$ \\
  \hline\hline
\end{tabular}
\label{table01}
\bigskip
\end{table}

\begin{figure*}[]
    \centering
    \includegraphics[width=0.45\textwidth,clip]{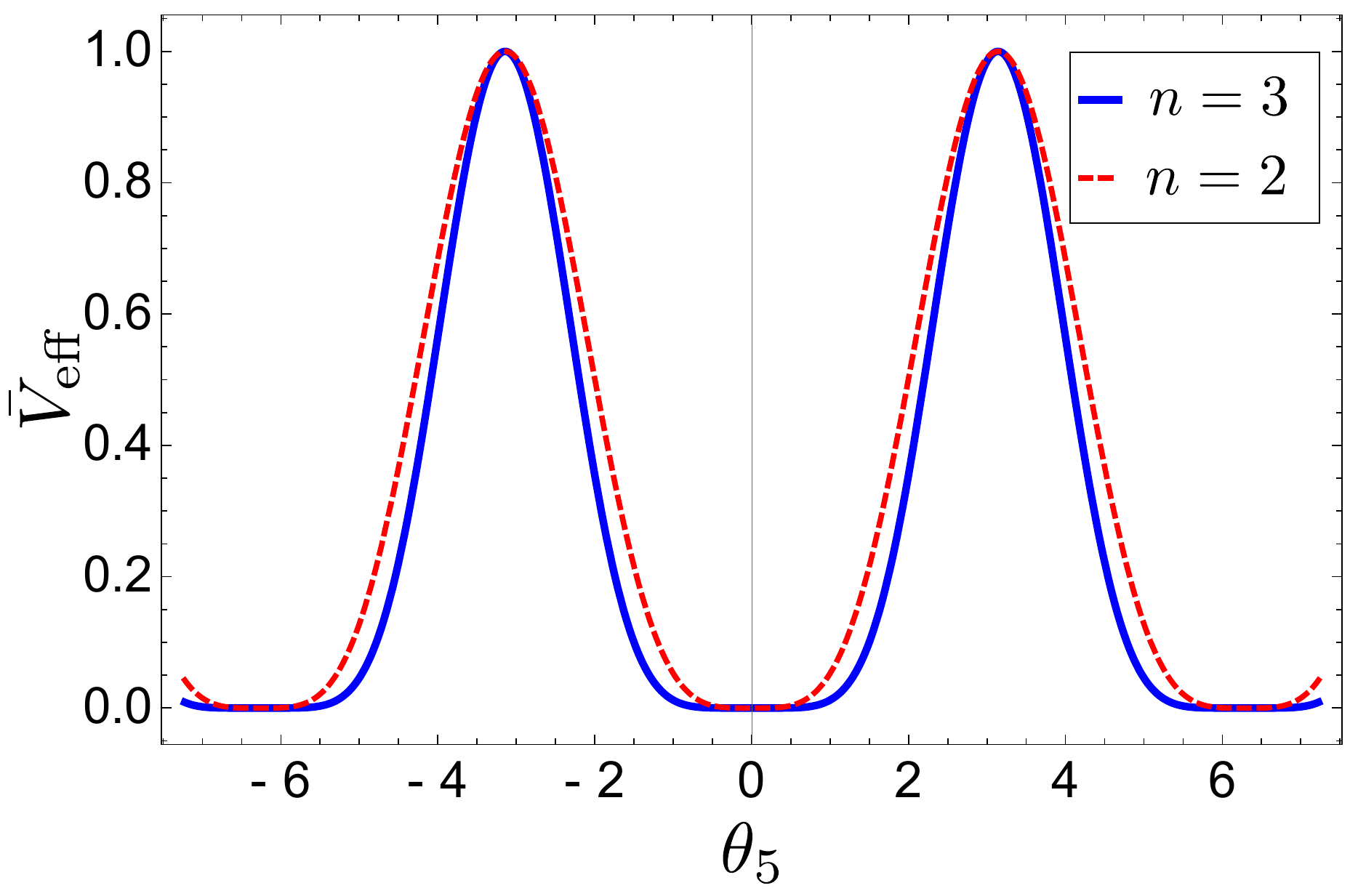}
    \hspace{0.03\linewidth}
    \mbox{\raisebox{1.1mm}{\includegraphics[width=0.458\textwidth,clip]{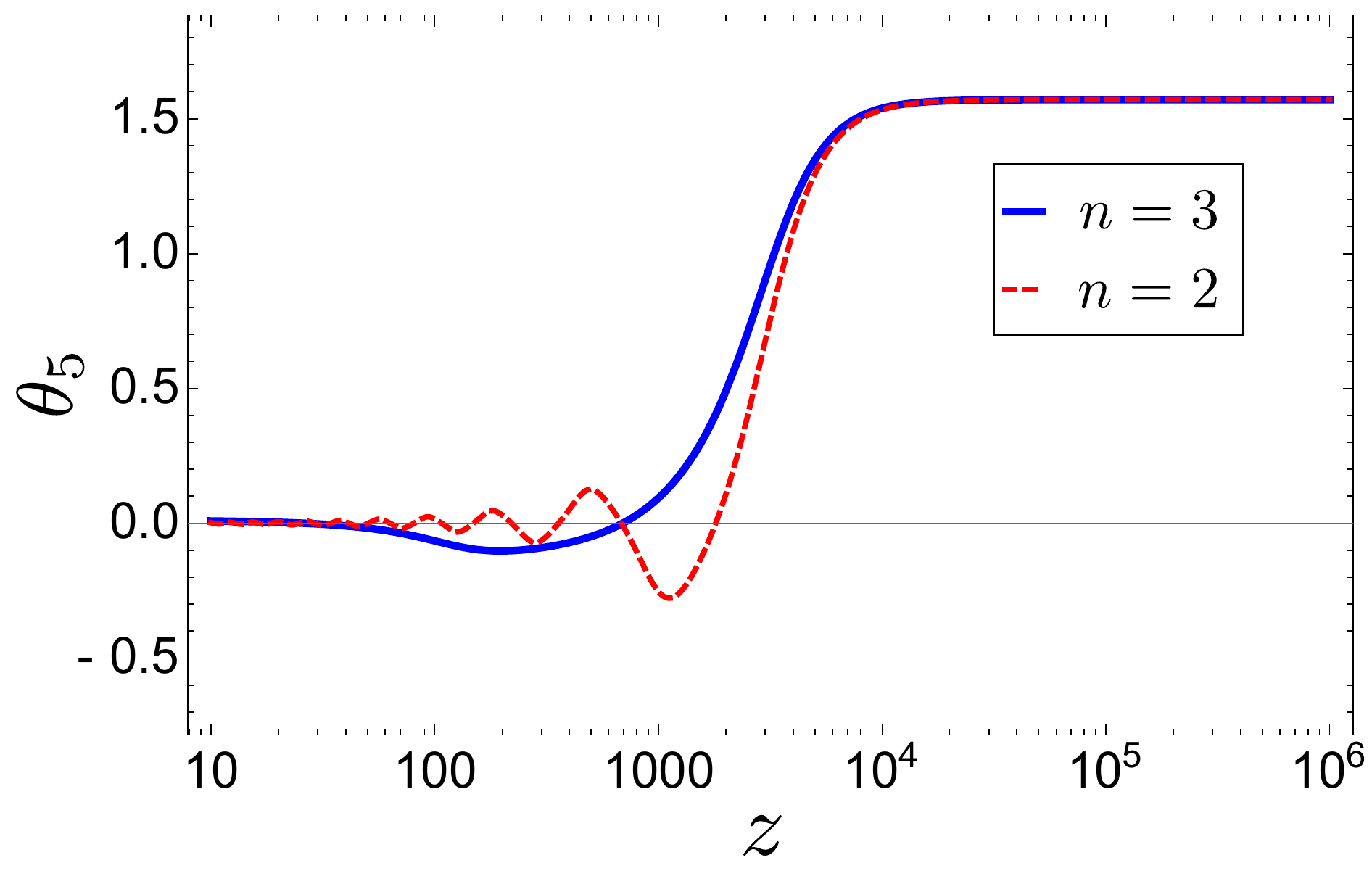}}}
    \caption{The effective potential for $\theta_5$ (left) and the time evolution of $\theta_5$ as functions of the redshift (right). 
    Here, the shown potential is normalized as $\bar V_{\rm eff}(\theta_5)=V_{\rm eff}(\theta_5)/V_{\rm eff}(\pi)$. The dashed (solid) line represents the $n=2$ ($n=3$) case.
    The left figure shows that the flatness of the potential is obtained around $\theta_5=0$, where the potential is approximately written as $V_{\rm eff}(\theta_5)\simeq V^{(n)}_{\rm eff}(\theta_5)\propto\theta_5^{2n}$.
    The right figure shows that $\theta_5$ is frozen in the potential for $z\gtrsim 4000$ due to the Hubble friction. 
    For $z\sim 4000$, $\theta_5$ begins fast-roll in its potential and finally oscillates around the potential minimum.
    }
    \bigskip
    \label{fig_pot}
\end{figure*}

On the left of Fig.~\ref{fig_pot}, we show the normalized effective potential $\bar V_{\rm eff}(\theta_5)$, whose maximal value is one. 
The dashed (solid) line represents the potential in the $n=2$ ($n=3$) model.
Around $\theta_5=0$, the potential is approximately written as the polynomial form $V_{\rm eff}(\theta_5)\simeq V^{(n)}_{\rm eff}(\theta_5)\propto \theta_5^{2n}$.
We see that the slope of the potential around the minima is much flatter in the $n=3$ case.  
On the right of Fig.~\ref{fig_pot}, we also show the time evolution's of $\theta_5$ in the $n=2$ and $n=3$ models by dashed and solid lines.
For $z\gtrsim 4000$, $\theta_5$ is frozen in its potential due to the Hubble friction.  For $z\sim 4000$, $\theta_5$ starts to roll down in its potential and then oscillates around $\theta_5=0$. 
We see that the oscillation period in the $n=3$ case is much longer than the one in the $n=2$ case, as expected from the potential forms.

\begin{figure*}[]
    \centering
    \includegraphics[width=0.45\textwidth,clip]{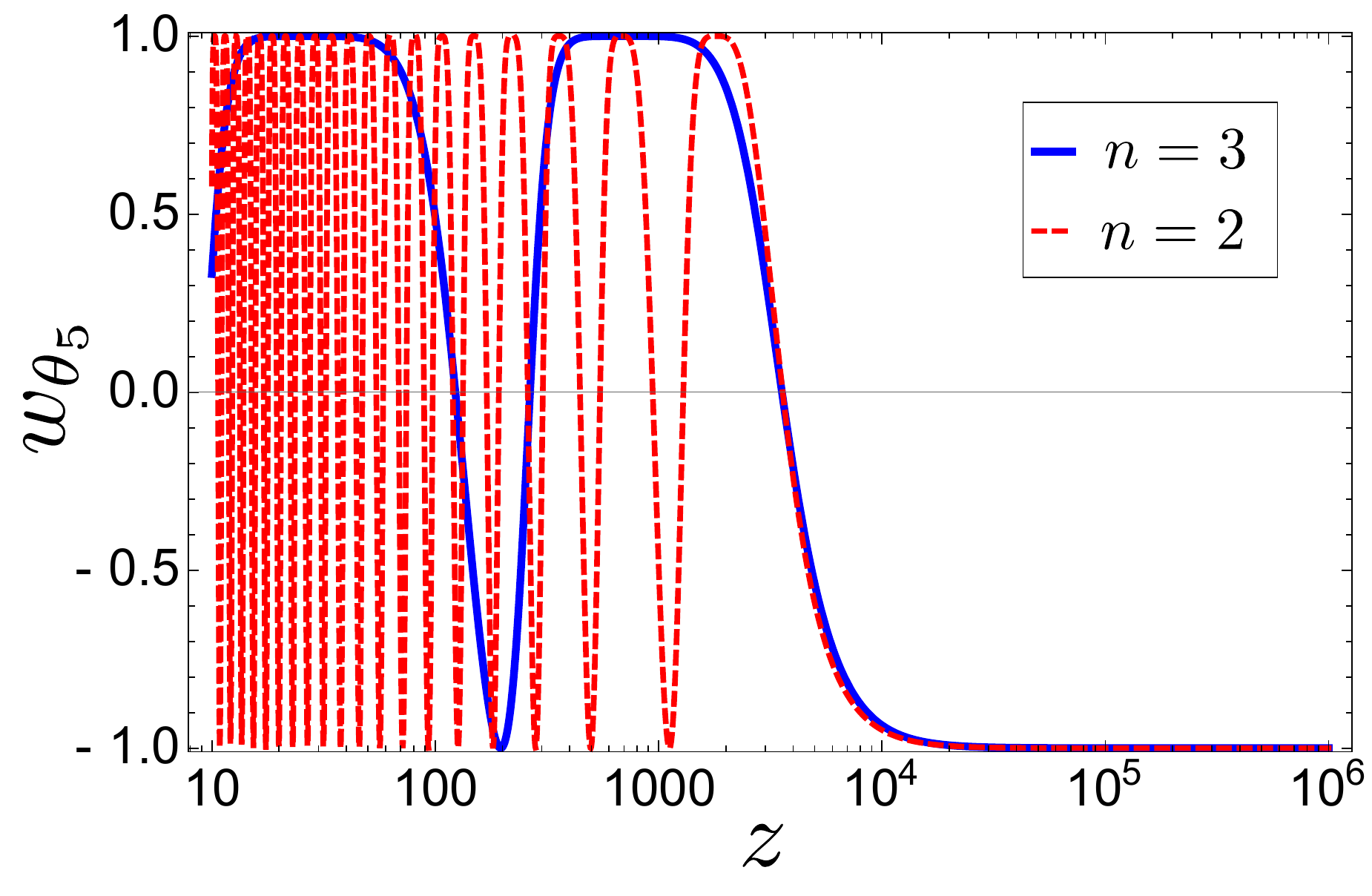}
    \hspace{0.04\linewidth}
    \includegraphics[width=0.45\textwidth,clip]{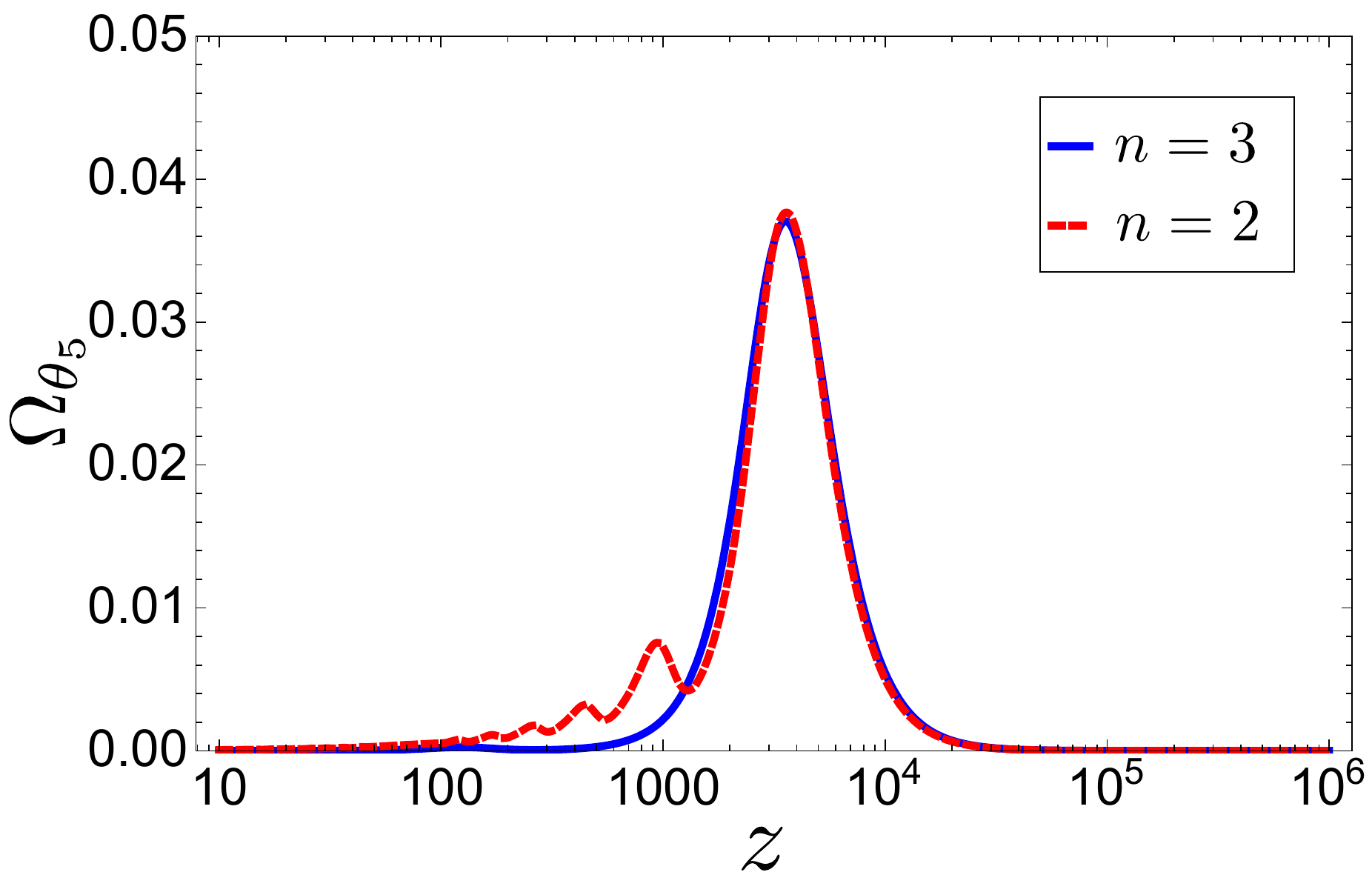}
    \caption{ The EoS $w_{\theta_5}$ (left) and the energy density parameter $\Omega_{\theta_5}$ (right) as functions of the redshift. 
    The dashed (solid) line represents the $n=2$ ($n=3$) case.
    For $z\gtrsim 4000$, $w_{\theta_5} \simeq -1$ holds like the dark energy. 
    When $\theta_5$ starts to oscillate, $w_{\theta_5}$ also oscillates such that the $w_{\theta_5}>0$ period is more extended than the $w_{\theta_5}<0$ one.
    Thus, the time-averaged value of $w_{\theta_5}$ satisfies $\langle w_{\theta_5} \rangle>0$.
    For $z>4000$, as $w_{\theta_5}\simeq -1$ is satisfied, $\Omega_{\theta_5}$ decays more slowly than the background matter energy density parameter.
    For $z<4000$, since the EoS takes $\langle w_{\theta_5} \rangle>0$, $\Omega_{\theta_5}$ proceeds to dilute.
    As seen from the figure, the dilution is more efficient for the $n=3$ case.}
\bigskip
  \label{fig_eos}
\end{figure*} 

In Fig.~\ref{fig_eos}, the EoS and the energy density parameter of the EDE phase are shown.
From the left in Fig.~\ref{fig_eos}, we see that the EoS $w_{\theta_5}$ also starts to oscillate when the EDE phase begins to oscillate around the minimum of the potential.  For the $n=3$ case, the EoS keeps $w_{\theta_5}\simeq 1$ relatively longer than in the $n=2$ case, as expected from the approximate formula in Eq.~\eqref{avephieos}.
From the right in Fig.~\ref{fig_eos}, we see that the EDE gives a few percent contribution to the total energy density of the universe around $z = 4000$.
Before $z\sim 4000$, $\rho_{\theta_5}$ decays more slowly than the background matter and radiation energy densities, as understood by $w_{\theta_5}<w_m<w_r$. 
Then, around $z = 4000$, the fast-roll of the EDE phase starts and triggers to change the EoS as $\langle w_{\theta_5} \rangle >0$. 
Thus, the EDE quickly dilutes after $z\sim 4000$.
For the $n=3$ case, dilution is much faster than the one in the $n=2$ case, as expected from the behavior of the EoS.

Let us discuss the parameter dependence of the EDE properties in our model.
The general discussion in Sec.~\ref{sec:pngb} is useful for understanding the parameter dependence.
From Eq.~\eqref{potmasscon1}, we see that $f\sim \delta^{1/2}M_{\rm P}$ holds, where $f$ and $\delta$ are defined in Eqs.~\eqref{pNGBV} and~\eqref{pot_scale_const1}. 
In our scenario, $\delta$ corresponds to a maximum value of $\Omega_{\rm \theta_5}(z)$, which we denote by $\Omega_{\rm max}$.
In addition, from the definition of the Wilson line phase, $\theta_5=\tilde g_DM_c^{-1}\vev{A_5^{(0)}}$, we realize $f\sim \tilde g_D^{-1}M_c$. 
Thus, we expect that a relation $\Omega_{\rm max}\sim \tilde g_D^{-2}M_c^2/M_{\rm P}^2$ is satisfied in our scenario.

\begin{figure*}[]
    \centering
    \includegraphics[width=0.45\textwidth,clip]{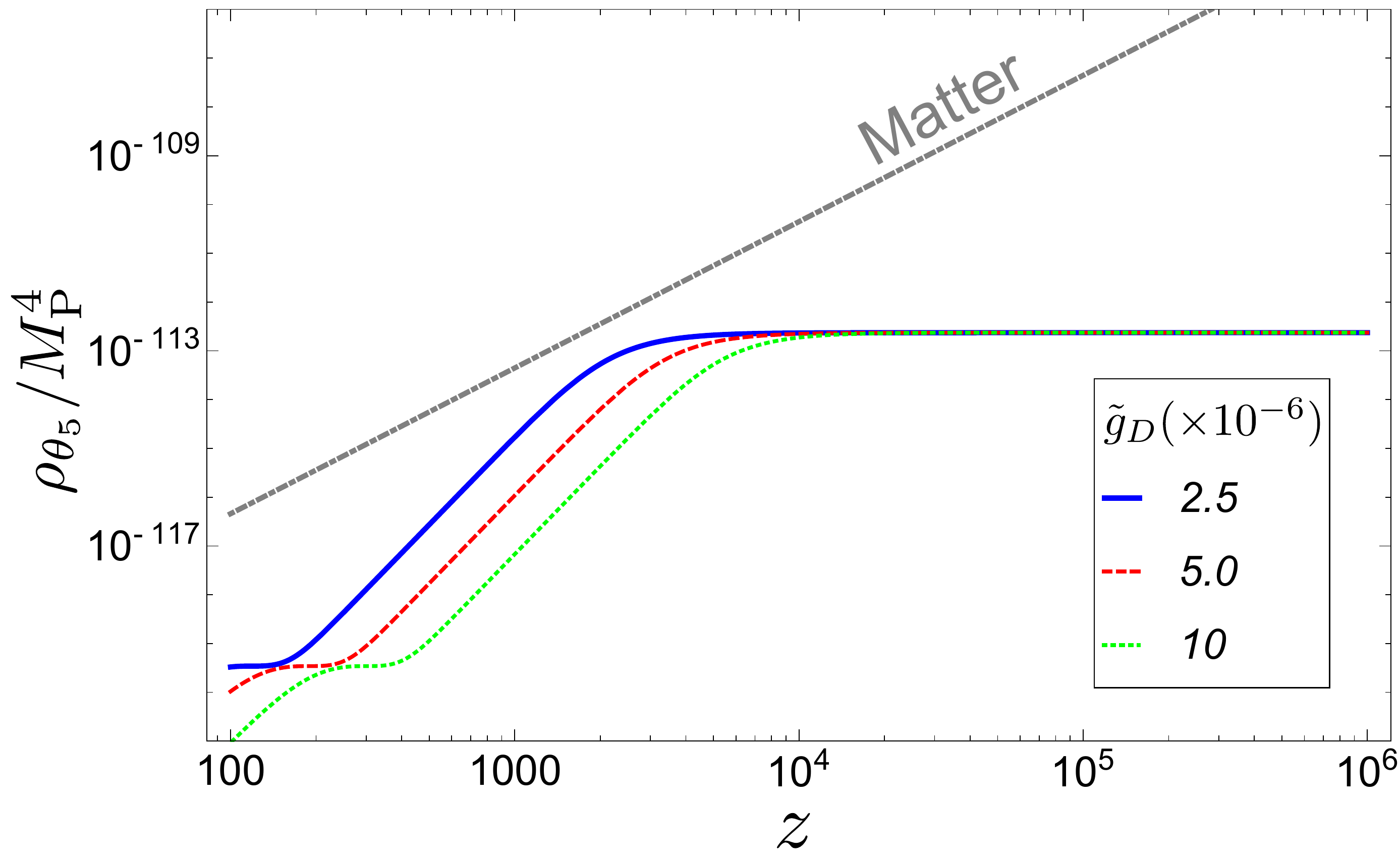}
    \hspace{0.04\linewidth}
    \includegraphics[width=0.45\textwidth,clip]{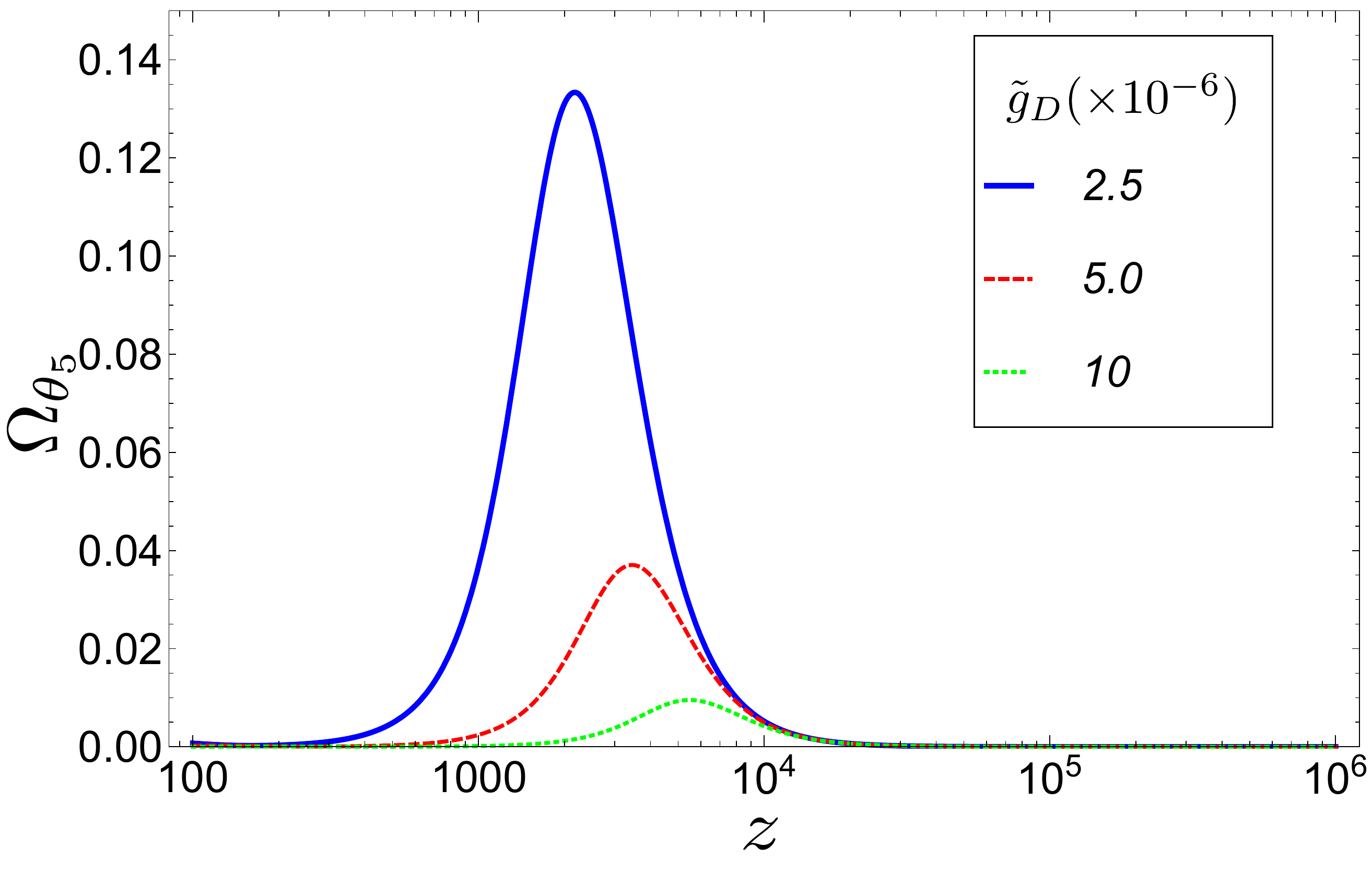}
    \caption{The $\tilde{g}_D$ dependence of the energy density $\rho_{\theta_5}$~(left) and the energy density parameter $\Omega_{\theta_5}$~(right). 
    We take the other parameters as the $n=3$ case in Table~\ref{table01}. 
    If $\tilde{g}_D$ is increased, the fast-roll of $\theta_5$ starts more earlier, and the peak value of the energy density, $\Omega_{\rm max}$, is suppressed.}
  \label{fig06}
\end{figure*}

To clarify the parameter dependence, we numerically study the $n=3$ model in more detail.
On the left in Fig.~\ref{fig06}, we show how the energy density $\rho_{\theta_5}$ depends on the coupling constant
$\tilde{g}_D$.
As explained, the equation of motion~\eqref{eomth51} depends on $\tilde{g}_D$, whereas the magnitude of the potential $V_{\rm eff}(\phi_5)$ does not.
Thus, the energy density before $z\sim 4000$ is unchanged even if we vary $\tilde{g}_D$.
On the other hand, $z_{\rm max}$, which we define by
$\Omega_{\rm max}= \Omega_{\theta_5}(z_{\rm max})$, depends on $\tilde{g}_D$.
If $\tilde{g}_D$ is increased, $z_{\rm max}$ is also
increased, and the fast-roll of $\theta_5$ starts earlier.  On the right in Fig.~\ref{fig06}, we show how the energy density parameter $\Omega_{\theta_5}$ depends on the coupling constant $\tilde{g}_D$.
Since the energy density of non-relativistic matter $\rho_m$ decays at a constant rate, if $\tilde{g}_D$ is increased, $\Omega_{\theta_5}(z_{\rm max})$ is decreased.

\begin{figure*}[]
    \centering
    \includegraphics[width=0.45\textwidth,clip]{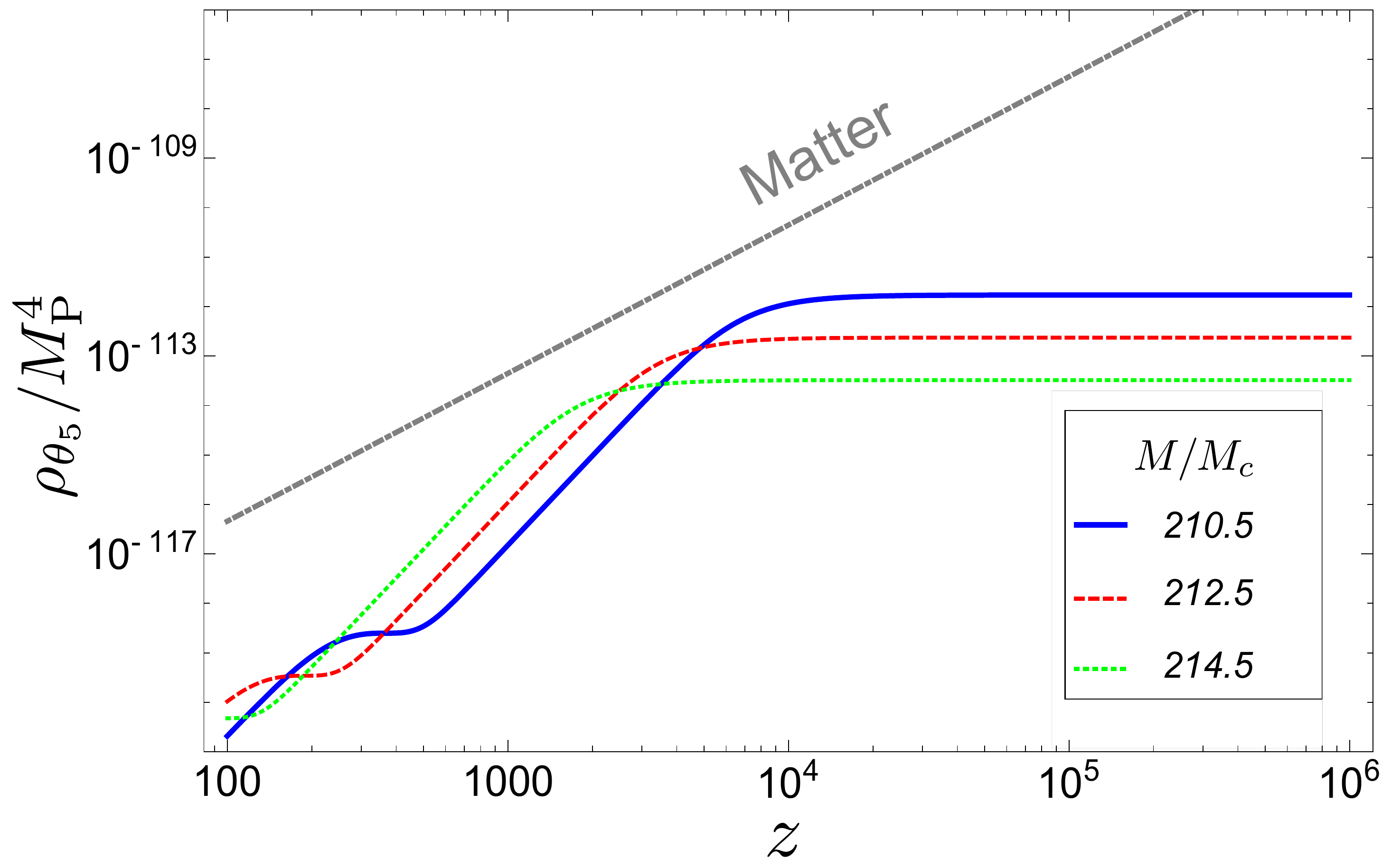}
    \hspace{0.04\linewidth}
    \includegraphics[width=0.45\textwidth,clip]{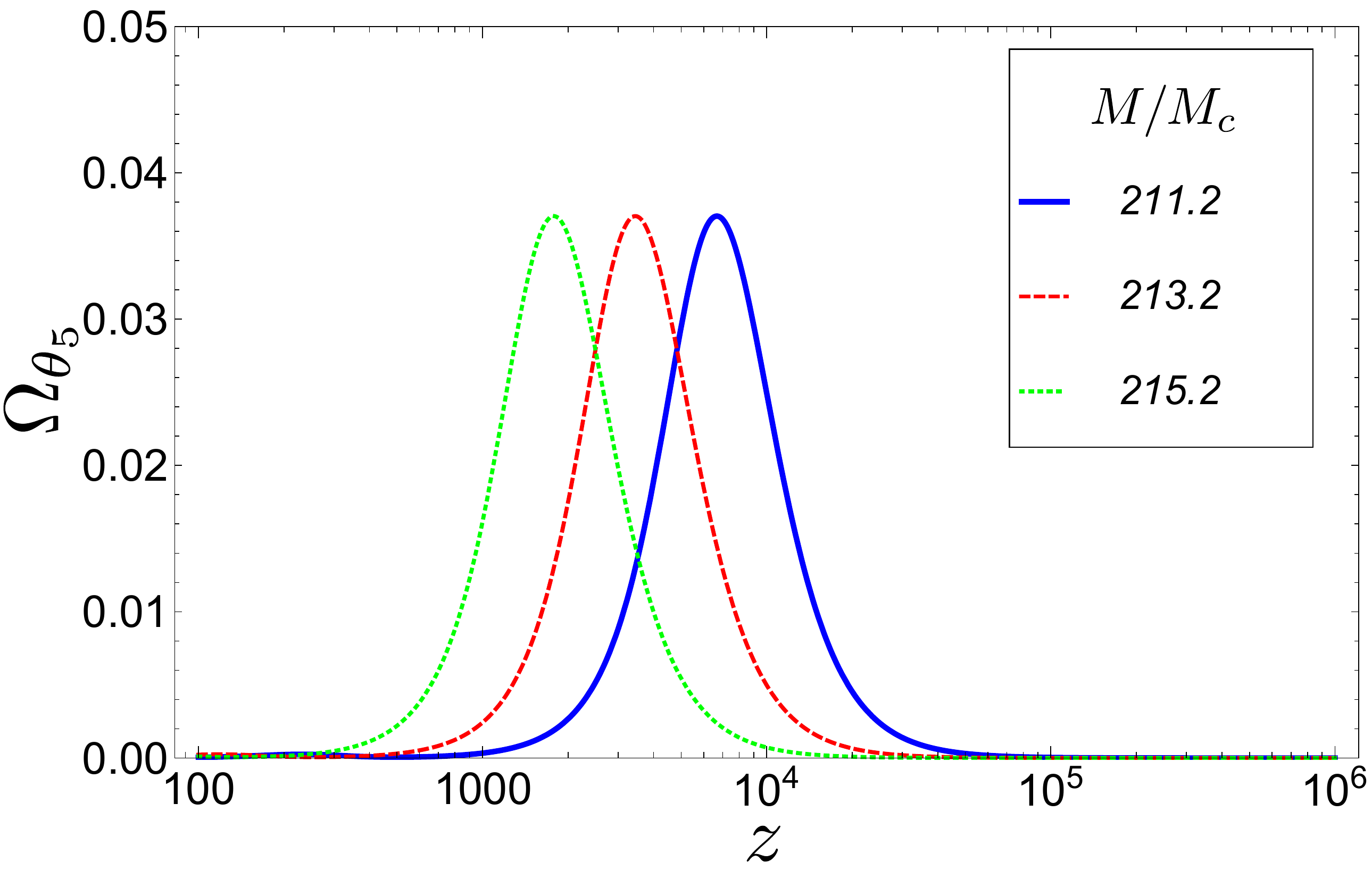}
    \caption{The bulk mass $M$ dependence of the energy density $\rho_{\theta_5}$~(left) and the energy density parameter $\Omega_{\theta_5}$~(right).
    We take the other parameters as the $n=3$ case in Table~\ref{table01}.
    If $M$ is increased, $z_{\rm max}$ is reduced, while $\Omega_{\rm max}$ is unchanged.}
  \label{fig07}
\end{figure*}

On the left and right in Fig.~\ref{fig07}, we show how
$\rho_{\theta_5}$ and $\Omega_{\theta_5}$ depend on the bulk mass parameter $M$, which affects the size of the potential
$V_{\rm eff}(\theta_5)$. 
From Eq.~\eqref{th5pot1}, we see that a larger $M$ suppresses the potential. 
As a result, if we increase $M$, both the energy density before $z\sim 4000$ and at $z_{\rm max}$ are decreased.
On the other hand, the maximum value of the energy density parameter is determined by $\tilde g_D$, $M_c$, and $M_{\rm P}$, as explained. 
Thus, $\Omega_{\rm max}$ does not depend on $M$ as shown in the figure.

\begin{figure}[]
  \centering
      \includegraphics[width=8cm,clip]{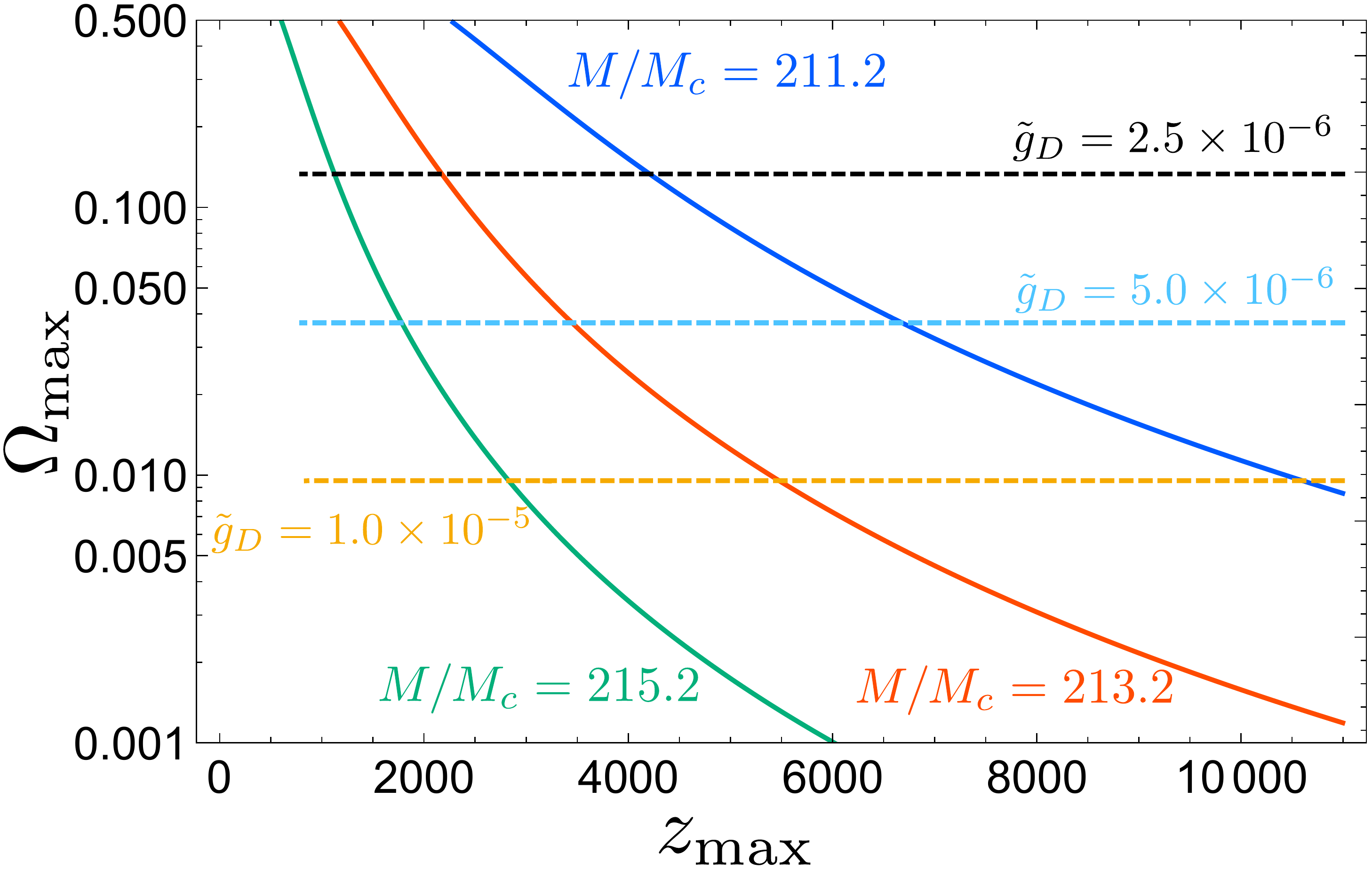}
      \caption{The predicted values of $z_{\rm max}$ and $\Omega_{\rm max}$ as functions of $\tilde{g}_D$ and $M/M_c$.
      The solid vertical lines represent the case with fixed bulk mass and the dashed horizontal lines represent the case with fixed coupling constants.
      The fixed values are shown next to each line.
      The three intersections of the cyan dashed line and solid lines correspond to the peaks appeared on the right in Fig.~\ref{fig07}, and the three intersections of the red solid line and dashed lines correspond to the peaks appeared on the right in Fig.~\ref{fig06}.}
  \bigskip
\label{fig08}
\end{figure}

In Fig.~\ref{fig08}, we show predicted values of $z_{\rm max}$ and $\Omega_{\rm max}$ as functions of $\tilde{g}_D$ and $M$. 
Suppressed values of the coupling constant and bulk mass increase both $z_{\rm max}$ and $\Omega_{\rm max}$. 
If precise constraints on $z_{\rm max}$ and $\Omega_{\rm max}$ are obtained from future measurements, we may find implications for the fundamental parameters $\tilde{g}_D$ and $M/M_c$ in our model by using the relation shown in Fig.~\ref{fig08}.
Note that, in a region where $\Omega_{\rm max}$ takes a relatively large value, the approximation used in the calculation may be worse because only $\rho_m$ is taken into account as the background energy for solving the Friedmann equation.
To obtain a more accurate result, a complete treatment of the Friedmann equation is required, which is left for future work.

We can estimate the $H_0$ value predicted in this scenario by evaluating the sound horizon.
Since the difference in $D(z_*)$ between $\Lambda$CDM and EDE model is tiny and negligible, the ratio of the Hubble constants of these models is given by
\begin{align}
	\frac{H_0^{\rm EDE}}{H_0^{\Lambda {\rm CDM}}}=\frac{r_{s,*}^{\Lambda {\rm CDM}}}{r_{s,*}^{\rm EDE}},
\end{align}
where $H_0^x$ and $r_{s,*}^x\ (x=\Lambda{\rm CDM}, {\rm EDE})$ denotes the Hubble constant and the sound horizon in each model.
In Sec.~\ref{sec:pngb}, we obtained $r_{s,*}^{\Lambda{\rm CDM}}=144\ {\rm Mpc}$ from Eq.~\eqref{eqn:dzandhorizon} and the $Planck$ results.
We can calculate  $r_{s,*}^{\rm EDE}$ by adding the energy density of $\theta_5$ to the energy density in Eq.~\eqref{eqn:homega}.
In the calculations of $r_{s,*}^{EDE}$, to include the effects from radiation properly, we use the relation

\begin{align}\label{eqn:alltz}
    t(z)
    =
    \frac{2}{3\bar{H}_0}\frac{1}{\Omega_{m,0}h^2}
    \left[
    \frac{1}{1+z}\sqrt{\Omega_{r,0}h^2+\frac{\Omega_{m,0}h^2}{1+z}}
    -
    \frac{2\Omega_{r,0}h^2}{\Omega_{m,0}h^2}
    \left(
    \sqrt{\Omega_{r,0}h^2+\frac{\Omega_{m,0}h^2}{1+z}}-\sqrt{\Omega_{r,0}h^2}
    \right)
    \right],
\end{align}
to rewrite $\rho_{\theta}$ as a function of $z$.
In addition, we assume $h=0.7$ to obtain $\rho_{\theta_5}h^2$.
Under these approximations, we get
\begin{align}
    \frac{H_0^{\rm EDE}}{H_0^{\Lambda {\rm CDM}}}\approx 1.05
\end{align}
for the $n=3$ model with $\tilde{g}_4=2.95\times 10^{-6}$ and $M/M_c=213.1$, which gives $(z_{\rm max},\Omega_{\rm max})\sim (2500,0.10)$.
Using the value $H_0^{\Lambda{\rm CDM}}=67.4\ {\rm km\ s^{-1} Mpc^{-1}}$ yields $H_0^{\rm EDE}=70.8\ {\rm km\ s^{-1}Mpc^{-1}}$.
If we perform detailed analysis including cosmological perturbations without the approximations used so far, we can obtain a more accurate $H_0$ value and other cosmological parameters, which are required to study the consistency with late time constrains.
In addition, we expect that our model can be distinguished from the other solutions to the Hubble tension thorough such analysis, which is left for future studies.

We have analyzed the time evolution of the EDE phase by using the leading effective potential in Eq.~\eqref{th5pot1}.
As a complement to
our analysis, we examine subleading corrections to the effective potential and their influence on our discussion. As stated in Sec.~\ref{sec:hdg}, the potential~\eqref{vtheta5w} is obtained in the one-loop approximation. 
Thus, there are two-loop effects to the leading potential~\eqref{th5pot1} as subleading corrections.
In addition, to get the leading potential~\eqref{th5pot1}, we discard $\ell\geq 2$ terms in Eq.~\eqref{vtheta5w}, which also give corrections.

In our scenario, EDE properties for $z>z_{\rm EDE}$ are mainly determined by the scale of the effective potential for the EDE phase.
Thus, higher-order corrections do not give significant effects for $z>z_{\rm EDE}$.
On the other hand, for $z<z_{\rm EDE}$, the time evolution of the EDE phase is sensitive to the structure of the potential.
For example, in the $n=2$ model, the leading potential is given by $V_{\rm eff}(\theta_5)\simeq V^{(2)}_{\rm eff}(\theta_5)\propto\theta_5^4$.
If the higher-order corrections include ${\cal O}(\theta_5^2)$ terms, the form of the potential near the minimum largely changes.
The ${\cal O}(\theta_5^2)$ contribution dominates the potential in the region where $|\theta_5|$ is smaller than a critical value $\theta_c$, which is determined by the size of the corrections.
Therefore, if $\theta_c$ is sufficiently suppressed, we expect that the time evolution of the system is not greatly affected by the corrections.

To estimate the influence of the higher-order corrections on our analysis, let us consider the case where the leading potential $V_{\rm eff}(\theta_5)\simeq V^{(n)}_{\rm eff}(\theta_5)\propto \theta_5^{2n}$ receives a correction $\delta V^{(m)}\sim {\cal O}(\theta^{2m})$ $(n>m\geq 0)$.
In this case, we can generally parametrize the potential as
\begin{align}
  V_{\rm eff}(\theta_5)+\delta V^{(m)}\propto \theta_5^{2n}+\epsilon \theta_5^{2m}
  =\theta_5^{2n}(1+\epsilon \theta_5^{-2(n-m)}), 
\end{align}
where $\epsilon$ gives the size of the correction. 
For a given $\epsilon$, the correction term grows for $|\theta_5| \lesssim \epsilon^{1/[2(n-m)]} \equiv \theta_C$ in the potential.
For a case with $\theta_C\ll 1$, the time evolution of $\theta_5$ is similar to the $\epsilon=0$ cases, and qualitative results are the same as the discussions given above.
We numerically check that the corrections of $\theta_C\lesssim 10^{-2}$ hardly change the results of our numerical analysis.

We can roughly estimate the magnitude of $\theta_C$ expected to arise in our model.
If a higher-order correction originates from two-loop effects, we expect $|\epsilon|\sim \tilde{g}_{D}^2/(4\pi^2)$, which is conservatively smaller than $10^{-10}$ in our numerical calculations with $M_c/M_{\rm P}=1.0\times 10^{-6}$.
Thus, we find $\theta_C\sim [\tilde{g}_{D}^2/(4\pi^2)]^{1/[2(n-m)]} \lesssim 10^{-10/[2(n-m)]}$.  To obtain $\theta_C\lesssim 10^{-2}$, the condition $10/[2(n-m)]>2$ should be satisfied.
This condition is naturally satisfied for the $n=2$ and $3$ cases, even for $m=1$.
On the other hand, for $n\geq 4$ cases, our scenario may be disturbed by two-loop corrections to the effective potential.
Let us remark that if a smaller value of $M_c/M_{\rm P}$ is chosen, a required value of $\tilde g_D^2$ also becomes smaller, and higher-order corrections tend to be suppressed.
In this case, two-loop corrections in $n\geq 4$ cases are expected to be neglected.
For higher-order corrections originating from $\ell\geq 2$ effects, we expect that they are strongly suppressed as $|\epsilon|\sim e^{-M_\varphi/M_c}\sim e^{-100}\sim 10^{-44}$, which cannot impact the results.

Finally, we comment on the effects of non-degenerate bulk masses.
The discussion given in this section is based on the case with degenerate bulk mass parameters.
If non-degeneracy exists, a bulk mass of a field $\varphi$ can be expressed by $M+\Delta M_\varphi$, where $M$ is a typical bulk mass scale.
For the case with $|\Delta M_\varphi/M|\ll 1$, we can approximately parametrize the effective potential such as in Eq.~\eqref{th5pot1} by substituting $n_\varphi$ for a parameter involving effects of non-zero $\Delta M_\varphi$. 
In this case, we can include non-degeneracy into our discussion by changing only the definition of $n_\varphi$.  Thus, with non-degeneracies, $n\geq 2$ models are possible, while some additional fine-tuning may be required.

%
\section{Discussions and conclusions}
\label{sec:conc}
%
The Hubble tension is the discrepancy that should be resolved by some extensions of the standard $\Lambda$CDM cosmology. 
We have shown that the 5D $U(1)_D$ gauge theory, introduced as a dark sector, provides suitable EDE to resolve the tension through the Wilson line phase dynamics.
The Wilson line phase associated with the zero-mode of $A_5$ has a radiatively induced PNGB type potential.
A tiny mass scale of the potential, which is suitable to explain EDE, is naturally obtained if the mass parameters in the 5D theory satisfy the relation in Eq.~\eqref{massrel}. 
In our scenario, $M_c$, the compactification scale of the extra dimension, is smaller than ${\cal O}(10^{-3}M_{\rm P})$.
We have examined the conditions for the matter contents to give the PNGB type potential that is well approximated by $2n$--th $(n\geq 2)$ polynomial potential around its minima.
We have demonstrated numerical calculations of the time evolution of the Wilson line phase in the $n=2$ and $3$ models.
The results show that the energy density of the Wilson line phase is a promising candidate for EDE.

More specific models of the 5D gauge theory may have additional implications for both cosmological and particle physics models.
Our scenario works in models with $S^1$ or $S^1/{\mathbb Z}_2$ compactification.
In the latter case, the 4D gauge field associated with $U(1)_D$ acquires a mass of the order of the compactification scale and is decoupled from low-energy physics.
On the other hand, for the $S^1$ case, the zero mode of 4D $U(1)_D$ gauge field remains light and can play some kind of cosmological role as a dark photon.
  For example, if the dark photon is a massless particle
  and does not have any interaction with the SM sector, it brings additional relativistic degrees of freedom and a larger value of $N_{\rm eff}$, which also has
been discussed as the solution to the Hubble
tension~\cite{Kreisch:2019yzn,DEramo:2018vss}.
Alternatively, the dark photon can have a mass and interactions with
the SM sector through additional mechanisms, such as Stueckelberg one
and a kinetic mixing to the photon. A massive dark photon can be a
candidate of dark matter, and the interactions with the SM sector can
predict phenomenologically and experimentally interesting
features~\cite{Fabbrichesi:2020wbt}.  
In addition, we have shown that the $n\geq 2$ models are explained by constrained $U(1)_D$ charges for degenerate bulk matter fields.
Such matter contents may be derived from more fundamental particle physics models.
  
In our scenario, the fundamental parameters in the 5D gauge theory, such as the $U(1)_D$ gauge coupling constant, the bulk mass parameters, and the compactification scale, are related to the property of EDE.
Thus, the future progress of cosmological measurements may give constraints and insights into particle physics models based on higher-dimensional gauge theories. 
To reveal the validity of our scenario, further investigations, such as a complete treatment of the Friedmann equation and analyses of cosmological perturbations, are required. 
We leave them as future work.

\bigskip \bigskip
\begin{center}
{\bf Acknowledgement}
\end{center}
The authors are grateful to Carolina Sayuri Takeda for helpful comments on the manuscript.
The work of Y.O. is supported in part by the Kyushu University Leading Human Resources Development Fellowship Program.
%
%
\newpage
\bigskip \bigskip
\appendix
\section{Derivation of the solution of Eq.~\eqref{nqconst1}}
\label{App:geneff}
%

In this appendix, we show that the solution of Eq.~\eqref{nqconst1} is given by Eq.~\eqref{sol_nqconst1}, which ensures that the effective potential for $\theta_5$ satisfies $V_{\rm eff}(\theta_5)\simeq V_{\rm eff}^{(n)}(\theta_5)$ around $\theta_5\simeq 0$ as discussed in Sec.~\ref{sec:u1gaugeede}.
To satisfy Eq.~\eqref{sol_nqconst1}, $U(1)$ charges and the degrees of freedom of bulk fields are constrained.

Suppose that there are $N$ kinds of bulk matter fields, which are scalars or fermions, with different charges in the 5D $U(1)$ gauge theory as studied in Sec.~\ref{sec:u1gaugeede}.
The effective potential for $\theta_5$ can be written as
\begin{gather}
    V_{\rm eff}(\theta_5)
	=
	-V_0
	\sum_{i=1}^{N}
	d_i (1-\cos(q_i \theta_5)),
	\label{App:effpot}
\end{gather}
where $d_i$ and $q_i$ represent the degrees of freedom and a $U(1)$ charge of a field $\varphi_i$.
We define $|q_i|\neq |q_j|$ for $i\neq j$ and $q_i\neq 0$.  
Here, we distinguish fields only for absolute values of their $U(1)$ charge.
Namely, if multiple fields have the same $U(1)$ charge squared, we denote them by the same label $i$.
The overall constant $V_0$ is given by Eq.~\eqref{th5pot1} and is irrelevant for the following discussion.

The potential~\eqref{App:effpot} takes an extreme value for $\theta_5=0$ and is expanded around $\theta_5=0$ as 
\begin{align}
  V_{\rm eff}(\theta_5)
  =&
  V_0
  \left[
  \frac{1}{2!}
  \left(- 
  \sum_{i=1}^N
  d_iq_i^2
  \right)
  \theta_5^2
  +
  \frac{1}{4!}
  \left(
  \sum_{i=1}^N
  d_i q_i^4
  \right)
  \theta_5^4
  +\cdots
  \right].
\end{align}
To obtain a potential dominated by the $\theta_5^{2n}$ ($n\geq 2$) term for $\theta_5\simeq 0$, the following $n-1$ conditions must be satisfied:
\begin{gather}
    \sum_{i=1}^N d_iq_i^{2k}=0, \qquad k=1,\dots,n-1, 
	\label{canccond}
\end{gather}
which can be interpreted as $n-1$ simultaneous equations for $d_i$.
While there is a trivial solution $d_i=0$, we don't concern it in the following.
If the above condition is satisfied, the effective potential is written as
\begin{align}
  V_{\rm eff}(\theta_5)
  =
  V_0
  \frac{(-1)^n}{(2n)!}
  \left( \sum_{i=1}^N
  d_iq_i^{2n}\right)
  \theta_5^{2n}
  +  {\cal O}(\theta_5^{2n+2}).
  \label{app:vefflead1}
\end{align}

For a given $q_i$, we can solve Eq.~\eqref{canccond} for $d_i$.
To see this, we remark that Eq.~\eqref{canccond} is not affected by a simultaneous rescaling of $d_i$ as $d_i\to cd_i$ with an arbitrary parameter $c$.
Thus, the set of parameters $\{d_1,d_2,\dots,d_N\}$ has $N-1$ degrees of freedom that are relevant to solve Eq.~\eqref{canccond}.
Therefore, to obtain a non-trivial solution of the $n-1$ conditions in Eq.~\eqref{canccond}, the condition $N-1\geq n-1$ should be satisfied.

In the following, we focus on the minimal case with $N=n$.
It is convenient to choose the field with $i=n$ as a basis of normalizations of $q_i$ and $d_i$ and define
\begin{align}
	\tilde{Q}_{i}\equiv \frac{q_i^2}{q_n^2},
	\qquad 
	\tilde{d}_{i}
	\equiv
	\frac{d_i}{d_n},
\end{align}
for $i=1,\cdots,n-1$.
Then, the conditions~(\ref{canccond}) can be expressed as
\begin{align}
	\label{ccondtilde}
  1+\sum_{j=1}^{n-1}
  \tilde{d}_j\tilde{Q}_j^{k}=0, \qquad k=1,\dots,n-1.
\end{align}
To solve these equations for $\tilde{d}_i$, let us introduce
\begin{align}\label{defV1}
\begin{aligned}
  V&=\begin{pmatrix}
      \tilde{Q}_1&\tilde{Q}_2&\cdots&\tilde{Q}_{n-1}
      \\
      \tilde{Q}_1^2&\tilde{Q}_2^2&\cdots&\tilde{Q}_{n-1}^2
      \\
      \vdots &\vdots& \ddots & \vdots
      \\
      \tilde{Q}_1^{n-1}&\tilde{Q}_2^{n-1}&\cdots&\tilde{Q}_{n-1}^{n-1}
	\end{pmatrix}
	\\
	&
	=
	\begin{pmatrix}
            1&1&\cdots&1
            \\
            \tilde{Q}_1&\tilde{Q}_2&\cdots&\tilde{Q}_{n-1}
            \\
            \vdots &\vdots& \ddots & \vdots
            \\
            \tilde{Q}_1^{n-2}&\tilde{Q}_2^{n-2}&\cdots&\tilde{Q}_{n-1}^{n-2}
	\end{pmatrix}
	\begin{pmatrix}
            \tilde{Q}_1&0&\cdots&0
            \\
            0&\tilde{Q}_2&\cdots&0
            \\
            \vdots &\vdots& \ddots & \vdots
            \\
            0&0&\cdots&\tilde{Q}_{n-1}
            \end{pmatrix},
\end{aligned}
\end{align}
where the first matrix on the right-hand side is known as the Vandermonde matrix. By using $V$, Eq.~\eqref{ccondtilde} is rewritten by
\begin{gather}
    V
    \begin{pmatrix}
		\tilde{d}_1
		\\
		\vdots
		\\
		\tilde{d}_{n-1}
            \end{pmatrix}
	= -
	\begin{pmatrix}
            1
            \\
            \vdots
            \\
            1
	\end{pmatrix}.
	\label{eq:SimEqforN}
\end{gather}
One sees that the determinant of $V$ is given by 
\begin{align}
	\det V=
	\left(
		\prod_{\substack{i,j=1\\ j> i}}^{n-1}
			(\tilde{Q}_j-\tilde{Q}_i)
	\right)
	\left(
		\prod_{m=1}^{n-1}
		\tilde{Q}_m
	\right).
\end{align}
Since $\tilde{Q}_j\neq \tilde{Q}_i$ for $i\neq j$ and $\tilde{Q}_i \neq 0$ hold in the present case, there exists the inverse of $V$, denoted by $V^{-1}$. Therefore, the solution is given by $\tilde d_i=-\sum_{k=1}^{n-1}(V^{-1})_{ik}$, where $(V^{-1})_{ik}$ is the $(i,k)$ element of $V^{-1}$.

To obtain an explicit expression for $\tilde d_i$, let us introduce
\begin{align}
	R_i(\tilde{Q})
	=
	\frac{\tilde{Q}}{\tilde{Q}_i}
	\prod_{\substack{k=1\\ k\neq i}}^{n-1}
	\frac{\tilde{Q}-\tilde{Q}_k}{\tilde{Q}_i-\tilde{Q}_k}.
\end{align}
By definition, $R_i(\tilde{Q})$ satisfies $R_i(\tilde{Q}_j)=\delta_{ij}$.
Here, let us remind that
$\sum_{k=1}^{n-1}(V^{-1})_{ik}(V)_{kj}=\sum_{k=1}^{n-1}(V^{-1})_{ik}\tilde Q^i_j= \delta_{ij}$ holds, where we have used the fact that the $(i,j)$ element of $V$ is given by $(V)_{ij}=\tilde Q^i_j$ as seen from Eq.~\eqref{defV1}.  
Thus, we find
\begin{align}
  R_i(\tilde{Q}_j)=\sum_{k=1}^{n-1}(V^{-1})_{ik}\tilde Q^k_j.
\end{align}
From the above discussions, $\tilde d_i$ is written as follows:
\begin{align}
  \tilde d_i=-\sum_{k=1}^{n-1}(V^{-1})_{ik}=-R_i(1)=
	-
	\frac{1}{\tilde{Q}_i}
	\prod_{\substack{k=1\\ k\neq i}}^{n-1}
	\frac{1-\tilde{Q}_k}{\tilde{Q}_i-\tilde{Q}_k}.
	\label{eqn:solni}
\end{align}
While we have chosen the $n$--th field as the basis of the normalizations, the above result should not depend on this choice.
As expected, one sees that the solution~\eqref{eqn:solni} is rearranged as
\begin{align}
  {d_i\over d_j}=-{q_j^2\over q_i^2}\prod_{\substack{k=1\\k\neq i,j}}^n{q_j^2-q_k^2\over q_i^2-q_k^2},
  \label{ap:ijsymform}
\end{align}
which holds for any $i,j\in\{1,\dots,n\}$ and $i\neq j$.

By using the above solution, we can rewrite the leading term of the effective potential in Eq.~\eqref{app:vefflead1}.
By using $\tilde{d}_i$ and $\tilde{Q}_i$, the coefficient appearing in Eq.~\eqref{app:vefflead1} is expressed as
\begin{align}
	\frac{(-1)^{n}}{(2n)!}
	\sum_{i=1}^n
	d_iq_i^{2n}
	=
	\frac{(-1)^n}{(2n)!}
	d_nq_n^{2n}
	\left(1+
	\sum_{i=1}^{n-1}
	\tilde{d}_i\tilde{Q}_i^{n}
  \right).
  \label{app:cofpot1}
\end{align}
Substituting Eq.~(\ref{eqn:solni}) into the last factor on the right-hand side of Eq.~\eqref{app:cofpot1}, we find
\begin{align}
    1+\sum_{i=1}^{n-1}\tilde{d}_i\tilde{Q}_i^{n}
    =
    1
    &
    -\sum_{i=1}^{n-1} \tilde{Q}_i^{{n-1}} \left[\prod_{\substack{k=1\\ k\neq i}}^{n-1} \frac{1-\tilde{Q}_k}{\tilde{Q}_i-\tilde{Q}_k}\right]
    \nonumber
    \\
    =
    1
    &
    -\tilde{Q}_1^{{n-1}}\prod_{\substack{k=1\\ k\neq 1}}^{n-1} \frac{1-\tilde{Q}_k}{\tilde{Q}_1-\tilde{Q}_k}
    \nonumber
    \\
    &
    -\tilde{Q}_2^{{n-1}}\prod_{\substack{k=1\\ k\neq 2}}^{n-1} \frac{1-\tilde{Q}_k}{\tilde{Q}_2-\tilde{Q}_k}
    \label{eqn:sqbra}
    \\
    &
    -\cdots-
    \tilde{Q}_{n-1}^{{n-1}}\prod_{\substack{k=1\\ k\neq {n-1}}}^{n-1} \frac{1-\tilde{Q}_k}{\tilde{Q}_{n-1}-\tilde{Q}_k}.
    \nonumber
\end{align}
Since the above quantity vanishes for $\tilde{Q}_i=1$ with an arbitrary $i$ and is linear for any $\tilde{Q}_i$, we can also write
\begin{align}
     1+
	\sum_{i=1}^{n-1}
	\tilde{d}_i\tilde{Q}_i^{n}
	=
	C(1-\tilde{Q}_1)(1-\tilde{Q}_2)\cdots(1-\tilde{Q}_{n-1}), 
	\label{eqn:anzatz}
\end{align}
where $C$ is a constant. One sees that $C=1$ should be satisfied by taking a specific value of $\tilde{Q}_i$, \eg, $\tilde{Q}_i=1-i$.
Thus, we get
\begin{align}
  \frac{(-1)^{n}}{(2n)!}
  \sum_{i=1}^n
  d_iq_i^{2n}
  &=
    \frac{(-1)^n}{(2n)!}
    d_nq_n^{2n}
    (1-\tilde{Q}_1)
    \cdots
    (1-\tilde{Q}_{n-1})
    \nonumber
  \\
  &=
    \frac{-1}{(2n)!}d_nq_n^{2}
    \left(q_1^2-q_n^2\right)
    \cdots
    \left(q_{n-1}^2-q_n^2\right)
    \nonumber
  \\
  &=\frac{-1}{(2n)!}d_{j}q_{j}^{2}\left(\prod_{\substack{k=1\\k\neq j}}^n(q_k^2-q_j^2)
  \right),
	\label{eqn:appnone}
\end{align}
where we have used Eq.~\eqref{ap:ijsymform} in the last equation.
Note that the last expression in Eq.~\eqref{eqn:appnone} holds for any $j\in\{1,\dots,n\}$.
Thus, the leading effective potential is written as Eq.~\eqref{approx_leadingpot1}.

\medskip


\begin{thebibliography}{99}
\bibitem{Planck:2018vyg}
N.~Aghanim \textit{et al.} [Planck],
``Planck 2018 results. VI. Cosmological parameters,''
Astron. Astrophys. \textbf{641}, A6 (2020)
[erratum: Astron. Astrophys. \textbf{652}, C4 (2021)]
[arXiv:1807.06209 [astro-ph.CO]].

\bibitem{Riess:2019cxk}
A.~G.~Riess, S.~Casertano, W.~Yuan, L.~M.~Macri and D.~Scolnic,
``Large Magellanic Cloud Cepheid Standards Provide a 1\% Foundation for the Determination of the Hubble Constant and Stronger Evidence for Physics beyond $\Lambda$CDM,''
Astrophys. J. \textbf{876}, no.1, 85 (2019)
[arXiv:1903.07603 [astro-ph.CO]].

\bibitem{Riess:2021jrx}
A.~G.~Riess, W.~Yuan, L.~M.~Macri, D.~Scolnic, D.~Brout, S.~Casertano, D.~O.~Jones, Y.~Murakami, L.~Breuval and T.~G.~Brink, \textit{et al.}
``A Comprehensive Measurement of the Local Value of the Hubble Constant with 1 km/s/Mpc Uncertainty from the Hubble Space Telescope and the SH0ES Team,''
[arXiv:2112.04510 [astro-ph.CO]].

\bibitem{DiValentino:2021izs}
E.~Di Valentino, O.~Mena, S.~Pan, L.~Visinelli, W.~Yang, A.~Melchiorri, D.~F.~Mota, A.~G.~Riess and J.~Silk,
``In the realm of the Hubble tension\textemdash{}a review of solutions,''
Class. Quant. Grav. \textbf{38}, no.15, 153001 (2021)
[arXiv:2103.01183 [astro-ph.CO]].

\bibitem{Anchordoqui:2015lqa}
L.~A.~Anchordoqui, V.~Barger, H.~Goldberg, X.~Huang, D.~Marfatia, L.~H.~M.~da Silva and T.~J.~Weiler,
``IceCube neutrinos, decaying dark matter, and the Hubble constant,''
Phys. Rev. D \textbf{92}, no.6, 061301 (2015)
[erratum: Phys. Rev. D \textbf{94}, no.6, 069901 (2016)]
[arXiv:1506.08788 [hep-ph]].

\bibitem{Ko:2017uyb}
P.~Ko, N.~Nagata and Y.~Tang,
``Hidden Charged Dark Matter and Chiral Dark Radiation,''
Phys. Lett. B \textbf{773}, 513-520 (2017)
[arXiv:1706.05605 [hep-ph]].

\bibitem{Buen-Abad:2018mas}
M.~A.~Buen-Abad, R.~Emami and M.~Schmaltz,
``Cannibal Dark Matter and Large Scale Structure,''
Phys. Rev. D \textbf{98}, no.8, 083517 (2018)
[arXiv:1803.08062 [hep-ph]].

\bibitem{Vattis:2019efj}
K.~Vattis, S.~M.~Koushiappas and A.~Loeb,
``Dark matter decaying in the late Universe can relieve the H0 tension,''
Phys. Rev. D \textbf{99}, no.12, 121302 (2019)
[arXiv:1903.06220 [astro-ph.CO]].

\bibitem{Pandey:2019plg}
K.~L.~Pandey, T.~Karwal and S.~Das,
``Alleviating the $H_0$ and $\sigma_8$ anomalies with a decaying dark matter model,''
JCAP \textbf{07}, 026 (2020)
[arXiv:1902.10636 [astro-ph.CO]].

\bibitem{Hryczuk:2020jhi}
A.~Hryczuk and K.~Jod\l{}owski,
``Self-interacting dark matter from late decays and the $H_0$ tension,''
Phys. Rev. D \textbf{102}, no.4, 043024 (2020)
[arXiv:2006.16139 [hep-ph]].

\bibitem{Kumar:2016zpg}
S.~Kumar and R.~C.~Nunes,
``Probing the interaction between dark matter and dark energy in the presence of massive neutrinos,''
Phys. Rev. D \textbf{94}, no.12, 123511 (2016)
[arXiv:1608.02454 [astro-ph.CO]].

\bibitem{DiValentino:2017iww}
E.~Di Valentino, A.~Melchiorri and O.~Mena,
``Can interacting dark energy solve the $H_0$ tension?,''
Phys. Rev. D \textbf{96}, no.4, 043503 (2017)
[arXiv:1704.08342 [astro-ph.CO]].

\bibitem{Banihashemi:2018has}
A.~Banihashemi, N.~Khosravi and A.~H.~Shirazi,
``Ginzburg-Landau Theory of Dark Energy: A Framework to Study Both Temporal and Spatial Cosmological Tensions Simultaneously,''
Phys. Rev. D \textbf{99}, no.8, 083509 (2019)
[arXiv:1810.11007 [astro-ph.CO]];
``Phase transition in the dark sector as a proposal to lessen cosmological tensions,''
Phys. Rev. D \textbf{101}, no.12, 123521 (2020)
[arXiv:1808.02472 [astro-ph.CO]].

\bibitem{Renk:2017rzu}
J.~Renk, M.~Zumalac\'arregui, F.~Montanari and A.~Barreira,
``Galileon gravity in light of ISW, CMB, BAO and H$_0$ data,''
JCAP \textbf{10}, 020 (2017)
[arXiv:1707.02263 [astro-ph.CO]].

\bibitem{Khosravi:2017hfi}
N.~Khosravi, S.~Baghram, N.~Afshordi and N.~Altamirano,
``$H_0$ tension as a hint for a transition in gravitational theory,''
Phys. Rev. D \textbf{99}, no.10, 103526 (2019)
[arXiv:1710.09366 [astro-ph.CO]].

\bibitem{Yan:2019gbw}
S.~F.~Yan, P.~Zhang, J.~W.~Chen, X.~Z.~Zhang, Y.~F.~Cai and E.~N.~Saridakis,
``Interpreting cosmological tensions from the effective field theory of torsional gravity,''
Phys. Rev. D \textbf{101}, no.12, 121301 (2020)
[arXiv:1909.06388 [astro-ph.CO]].

\bibitem{Odintsov:2020qzd}
S.~D.~Odintsov, D.~S\'aez-Chill\'on G\'omez and G.~S.~Sharov,
``Analyzing the $H_0$ tension in $F(R)$ gravity models,''
Nucl. Phys. B \textbf{966}, 115377 (2021)
[arXiv:2011.03957 [gr-qc]].

\bibitem{Dainotti:2022bzg}
M.~G.~Dainotti, B.~De Simone, T.~Schiavone, G.~Montani, E.~Rinaldi, G.~Lambiase, M.~Bogdan and S.~Ugale,
``On the Evolution of the Hubble Constant with the SNe Ia Pantheon Sample and Baryon Acoustic Oscillations: A Feasibility Study for GRB-Cosmology in 2030,''
Galaxies \textbf{10}, no.1, 24 (2022)
[arXiv:2201.09848 [astro-ph.CO]].

\bibitem{Vagnozzi:2019ezj}
S.~Vagnozzi,
``New physics in light of the $H_0$ tension: An alternative view,''
Phys. Rev. D \textbf{102}, no.2, 023518 (2020)
[arXiv:1907.07569 [astro-ph.CO]].

\bibitem{Alestas:2020mvb}
G.~Alestas, L.~Kazantzidis and L.~Perivolaropoulos,
``$H_0$ tension, phantom dark energy, and cosmological parameter degeneracies,''
Phys. Rev. D \textbf{101}, no.12, 123516 (2020)
[arXiv:2004.08363 [astro-ph.CO]].

\bibitem{DiValentino:2016hlg}
E.~Di Valentino, A.~Melchiorri and J.~Silk,
``Reconciling Planck with the local value of $H_0$ in extended parameter space,''
Phys. Lett. B \textbf{761}, 242-246 (2016)
[arXiv:1606.00634 [astro-ph.CO]].

\bibitem{DiValentino:2017zyq}
E.~Di Valentino, A.~Melchiorri, E.~V.~Linder and J.~Silk,
``Constraining Dark Energy Dynamics in Extended Parameter Space,''
Phys. Rev. D \textbf{96}, no.2, 023523 (2017)
[arXiv:1704.00762 [astro-ph.CO]].

\bibitem{Vagnozzi:2018jhn}
S.~Vagnozzi, S.~Dhawan, M.~Gerbino, K.~Freese, A.~Goobar and O.~Mena,
``Constraints on the sum of the neutrino masses in dynamical dark energy models with $w(z) \geq -1$ are tighter than those obtained in $\Lambda$CDM,''
Phys. Rev. D \textbf{98}, no.8, 083501 (2018)
[arXiv:1801.08553 [astro-ph.CO]].

\bibitem{Yang:2018qmz}
W.~Yang, S.~Pan, E.~Di Valentino, E.~N.~Saridakis and S.~Chakraborty,
``Observational constraints on one-parameter dynamical dark-energy parametrizations and the $H_0$ tension,''
Phys. Rev. D \textbf{99}, no.4, 043543 (2019)
[arXiv:1810.05141 [astro-ph.CO]].

\bibitem{DiValentino:2019jae}
E.~Di Valentino, A.~Melchiorri, O.~Mena and S.~Vagnozzi,
``Nonminimal dark sector physics and cosmological tensions,''
Phys. Rev. D \textbf{101}, no.6, 063502 (2020)
[arXiv:1910.09853 [astro-ph.CO]].


\bibitem{DiValentino:2020naf}
E.~Di Valentino, A.~Mukherjee and A.~A.~Sen,
``Dark Energy with Phantom Crossing and the $H_0$ Tension,''
Entropy \textbf{23}, no.4, 404 (2021)
[arXiv:2005.12587 [astro-ph.CO]].

\bibitem{Dainotti:2021pqg}
M.~G.~Dainotti, B.~De Simone, T.~Schiavone, G.~Montani, E.~Rinaldi and G.~Lambiase,
``On the Hubble constant tension in the SNe Ia Pantheon sample,''
Astrophys. J. \textbf{912}, no.2, 150 (2021)
[arXiv:2103.02117 [astro-ph.CO]].

\bibitem{Karwal:2016vyq}
T.~Karwal and M.~Kamionkowski,
``Dark energy at early times, the Hubble parameter, and the string axiverse,''
Phys. Rev. D \textbf{94}, no.10, 103523 (2016)
[arXiv:1608.01309 [astro-ph.CO]].

\bibitem{Mortsell:2018mfj}
E.~M\"ortsell and S.~Dhawan,
``Does the Hubble constant tension call for new physics?,''
JCAP \textbf{09}, 025 (2018)
[arXiv:1801.07260 [astro-ph.CO]].

\bibitem{Poulin:2018cxd}
V.~Poulin, T.~L.~Smith, T.~Karwal and M.~Kamionkowski,
``Early Dark Energy Can Resolve The Hubble Tension,''
Phys. Rev. Lett. \textbf{122}, no.22, 221301 (2019)
[arXiv:1811.04083 [astro-ph.CO]].

\bibitem{Smith:2019ihp}
T.~L.~Smith, V.~Poulin and M.~A.~Amin,
``Oscillating scalar fields and the Hubble tension: a resolution with novel signatures,''
Phys. Rev. D \textbf{101}, no.6, 063523 (2020)
[arXiv:1908.06995 [astro-ph.CO]].

\bibitem{Hill:2020osr}
J.~C.~Hill, E.~McDonough, M.~W.~Toomey and S.~Alexander,
``Early dark energy does not restore cosmological concordance,''
Phys. Rev. D \textbf{102}, no.4, 043507 (2020)
[arXiv:2003.07355 [astro-ph.CO]].

\bibitem{Kamionkowski:2014zda}
M.~Kamionkowski, J.~Pradler and D.~G.~E.~Walker,
``Dark energy from the string axiverse,''
Phys. Rev. Lett. \textbf{113}, no.25, 251302 (2014)
[arXiv:1409.0549 [hep-ph]].

\bibitem{Poulin:2018dzj}
V.~Poulin, T.~L.~Smith, D.~Grin, T.~Karwal and M.~Kamionkowski,
``Cosmological implications of ultralight axionlike fields,''
Phys. Rev. D \textbf{98}, no.8, 083525 (2018)
[arXiv:1806.10608 [astro-ph.CO]].

\bibitem{Agrawal:2019lmo}
P.~Agrawal, F.~Y.~Cyr-Racine, D.~Pinner and L.~Randall,
``Rock 'n' Roll Solutions to the Hubble Tension,''
[arXiv:1904.01016 [astro-ph.CO]].

\bibitem{Alexander:2019rsc}
S.~Alexander and E.~McDonough,
``Axion-Dilaton Destabilization and the Hubble Tension,''
Phys. Lett. B \textbf{797}, 134830 (2019)
[arXiv:1904.08912 [astro-ph.CO]].

\bibitem{Niedermann:2019olb}
F.~Niedermann and M.~S.~Sloth,
``New early dark energy,''
Phys. Rev. D \textbf{103}, no.4, L041303 (2021)
[arXiv:1910.10739 [astro-ph.CO]].

\bibitem{Berghaus:2019cls}
K.~V.~Berghaus and T.~Karwal,
``Thermal Friction as a Solution to the Hubble Tension,''
Phys. Rev. D \textbf{101}, no.8, 083537 (2020)
[arXiv:1911.06281 [astro-ph.CO]].

\bibitem{Sakstein:2019fmf}
J.~Sakstein and M.~Trodden,
``Early Dark Energy from Massive Neutrinos as a Natural Resolution of the Hubble Tension,''
Phys. Rev. Lett. \textbf{124}, no.16, 161301 (2020)
[arXiv:1911.11760 [astro-ph.CO]].

\bibitem{Ye:2020btb}
G.~Ye and Y.~S.~Piao,
``Is the Hubble tension a hint of AdS phase around recombination?,''
Phys. Rev. D \textbf{101}, no.8, 083507 (2020)
[arXiv:2001.02451 [astro-ph.CO]].

\bibitem{Chudaykin:2020acu}
A.~Chudaykin, D.~Gorbunov and N.~Nedelko,
``Combined analysis of Planck and SPTPol data favors the early dark energy models,''
JCAP \textbf{08}, 013 (2020)
[arXiv:2004.13046 [astro-ph.CO]].

\bibitem{Haridasu:2020pms}
B.~S.~Haridasu, M.~Viel and N.~Vittorio,
``Sources of $H_0$-tension in dark energy scenarios,''
Phys. Rev. D \textbf{103}, no.6, 063539 (2021)
[arXiv:2012.10324 [astro-ph.CO]].

\bibitem{Vagnozzi:2021gjh}
S.~Vagnozzi,
``Consistency tests of \ensuremath{\Lambda}CDM from the early integrated Sachs-Wolfe effect: Implications for early-time new physics and the Hubble tension,''
Phys. Rev. D \textbf{104}, no.6, 063524 (2021)
[arXiv:2105.10425 [astro-ph.CO]].

\bibitem{Berghaus:2022cwf}
K.~V.~Berghaus and T.~Karwal,
``Thermal Friction as a Solution to the Hubble and Large-Scale Structure Tensions,''
[arXiv:2204.09133 [astro-ph.CO]].


\bibitem{Linde:2007fr}
For a review, see \eg, A.~D.~Linde,
``Inflationary Cosmology,''
Lect. Notes Phys. \textbf{738}, 1-54 (2008)
[arXiv:0705.0164 [hep-th]].

\bibitem{Tsujikawa:2013fta}
For a review, see \eg, S.~Tsujikawa,,
``Quintessence: A Review,''
Class. Quant. Grav. \textbf{30}, 214003 (2013)
[arXiv:1304.1961 [gr-qc]].

\bibitem{Turner:1983he}
M.~S.~Turner,
``Coherent Scalar Field Oscillations in an Expanding Universe,''
Phys. Rev. D \textbf{28}, 1243 (1983).

\bibitem{Arkani-Hamed:2003xts}
N.~Arkani-Hamed, H.~C.~Cheng, P.~Creminelli and L.~Randall,
``Extra natural inflation,''
Phys. Rev. Lett. \textbf{90}, 221302 (2003)
[arXiv:hep-th/0301218 [hep-th]];
``Pseudonatural inflation,''
JCAP \textbf{07}, 003 (2003)
[arXiv:hep-th/0302034 [hep-th]].

\bibitem{Pilo:2003gu}
L.~Pilo, D.~A.~J.~Rayner and A.~Riotto,
``Gauge quintessence,''
Phys. Rev. D \textbf{68}, 043503 (2003)
[arXiv:hep-ph/0302087 [hep-ph]].

\bibitem{Mather:1998gm}
J.~C.~Mather, D.~J.~Fixsen, R.~A.~Shafer, C.~Mosier and D.~T.~Wilkinson,
``Calibrator design for the COBE far infrared absolute spectrophotometer (FIRAS),''
Astrophys. J. \textbf{512}, 511-520 (1999)
[arXiv:astro-ph/9810373 [astro-ph]].

\bibitem{Lesgourgues:2006nd}
J.~Lesgourgues and S.~Pastor,
``Massive neutrinos and cosmology,''
Phys. Rept. \textbf{429}, 307-379 (2006)
[arXiv:astro-ph/0603494 [astro-ph]].

\bibitem{Akita:2020szl}
K.~Akita and M.~Yamaguchi,
``A precision calculation of relic neutrino decoupling,''
JCAP \textbf{08}, 012 (2020)
[arXiv:2005.07047 [hep-ph]].

\bibitem{Dimopoulos:2005ac}
S.~Dimopoulos, S.~Kachru, J.~McGreevy and J.~G.~Wacker,
``N-flation,''
JCAP \textbf{08}, 003 (2008)
[arXiv:hep-th/0507205 [hep-th]].

\bibitem{Marsh:2015xka}
D.~J.~E.~Marsh,
``Axion Cosmology,''
Phys. Rept. \textbf{643}, 1-79 (2016)
[arXiv:1510.07633 [astro-ph.CO]].

\bibitem{Riess:2016jrr}
A.~G.~Riess, L.~M.~Macri, S.~L.~Hoffmann, D.~Scolnic, S.~Casertano, A.~V.~Filippenko, B.~E.~Tucker, M.~J.~Reid, D.~O.~Jones and J.~M.~Silverman, \textit{et al.}
``A 2.4\% Determination of the Local Value of the Hubble Constant,''
Astrophys. J. \textbf{826}, no.1, 56 (2016)
[arXiv:1604.01424 [astro-ph.CO]].

\bibitem{DEramo:2018vss}
F.~D'Eramo, R.~Z.~Ferreira, A.~Notari and J.~L.~Bernal,
``Hot Axions and the $H_0$ tension,''
JCAP \textbf{11}, 014 (2018)
[arXiv:1808.07430 [hep-ph]].

\bibitem{Kreisch:2019yzn}
C.~D.~Kreisch, F.~Y.~Cyr-Racine and O.~Dor\'e,
``Neutrino puzzle: Anomalies, interactions, and cosmological tensions,''
Phys. Rev. D \textbf{101}, no.12, 123505 (2020)
[arXiv:1902.00534 [astro-ph.CO]].

\bibitem{Hebecker:2001jb}
A.~Hebecker and J.~March-Russell,
``The structure of GUT breaking by orbifolding,''
Nucl. Phys. B \textbf{625}, 128-150 (2002)
[arXiv:hep-ph/0107039 [hep-ph]].

\bibitem{Haba:2008ar}
N.~Haba, Y.~Kawamura and K.~Oda,
``Dynamical Rearrangement of Theta Parameter in Presence of Mixed Chern-Simons Term,''
Phys. Rev. D \textbf{78}, 085021 (2008)
[arXiv:0803.4380 [hep-ph]].

\bibitem{Kawamura:2011re}
Y.~Kawamura and T.~Miura,
``$\mu$ Parameter from Dynamical Rearrangement of U(1) and $\theta$ Parameter,''
Int. J. Mod. Phys. A \textbf{27}, 1250023 (2012)
[arXiv:1108.1004 [hep-ph]].

\bibitem{Abe:2016tfq}
Y.~Abe, Y.~Goto, Y.~Kawamura and Y.~Nishikawa,
``Conjugate boundary condition, hidden particles, and gauge-Higgs inflation,''
Mod. Phys. Lett. A \textbf{31}, no.35, 1650208 (2016)
[arXiv:1608.06393 [hep-ph]].

\bibitem{Delgado:1998qr}
A.~Delgado, A.~Pomarol and M.~Quiros,
``Supersymmetry and electroweak breaking from extra dimensions at the TeV scale,''
Phys. Rev. D \textbf{60}, 095008 (1999)
[arXiv:hep-ph/9812489 [hep-ph]].

\bibitem{Maru:2006ej}
N.~Maru and K.~Takenaga,
``Effects of bulk mass in gauge-Higgs unification,''
Phys. Lett. B \textbf{637}, 287-294 (2006)
[arXiv:hep-ph/0602149 [hep-ph]].

\bibitem{Fabbrichesi:2020wbt}
M.~Fabbrichesi, E.~Gabrielli and G.~Lanfranchi,
``The Dark Photon,''
[arXiv:2005.01515 [hep-ph]].

\end{thebibliography}
\end{document}